\documentclass{aa}
\usepackage[dvips]{graphicx}
\usepackage{natbib}
\usepackage{txfonts}
\usepackage{color}

\begin{document}
   \title{VLTI/AMBER
spectro-interferometric imaging of VX Sgr's inhomogenous outer
atmosphere\thanks{Based on the observations made with VLTI-ESO Paranal, Chile under the programme IDs 081.D-0005(A, B, C, D, E, F, G, H)}}
\titlerunning{}

   \author{A. Chiavassa
          \inst{1,2}
          \and
          S. Lacour \inst{3}
          \and
          F. Millour \inst{4}
          \and 
          T. Driebe \inst{4}
          \and
          M. Wittkowski \inst {5}
          \and
          B. Plez\inst{2}
          \and
          E. Thi\'ebeaut \inst {6}
         	\and
	E. Josselin \inst{2}
          \and
	B. Freytag\inst{7,8}
	\and
	M. Scholz \inst {9,10}
	\and
	X. Haubois \inst {3}
          }
  
   \offprints{A. Chiavassa}
    \institute{Max-Planck-Institut f\"{u}r Astrophysik, Karl-Schwarzschild-Str. 1, Postfach 1317, DÐ85741 Garching b. M\"{u}nchen, Germany\\
              \email{chiavass@mpa-garching.mpg.de}
            \and
            GRAAL, Universit\'{e} de Montpellier II - IPM, CNRS, Place Eug\'{e}ne Bataillon
	34095 Montpellier Cedex 05, France
         \and
         Observatoire de Paris, LESIA, CNRS/UMR 8109, 92190 Meudon, France
         \and
         Max-Planck-Institut f\"{u}r Radioastronomie, Auf dem H\"{u}gel 69, 53121 Bonn, Germany
         \and
         ESO, Karl-Schwarzschild-Str. 2, 85748 Garching, Germany
         \and
         AIRI/Observatoire de Lyon, France and Jean-Marie Mariotti Center, France
         \and
          Centre de Recherche Astrophysique de Lyon,
          UMR 5574: CNRS, Universit\'e de Lyon,
          \'Ecole Normale Sup\'erieure de Lyon,
          46 all\'ee d'Italie, F-69364 Lyon Cedex 07, France
            \and
          Department of Physics and Astronomy,
          Division of Astronomy and Space Physics,
          Uppsala University,
          Box 515, S-751~ 20 Uppsala,
          Sweden
          \and
	  Zentrum f\"{u}r Astronomie der Universität Heidelberg (ZAH), Institut f\"{u}r Theoretische Astrophysik, Albert Ueberle-Str. 2, 69120 Heidelberg, Germany  	
          \and
	  Sydney Institute for Astronomy, School of Physics, University of Sydney, Sydney, NSW 2006, Australia  
	     }

   \date{Received; accepted }

  \abstract
   {}
   {We aim to explore the photosphere of the very cool late-type
     star VX Sgr and in particular the existence and characterization of molecular layers above the continuum forming photosphere.}
   {We obtained interferometric observations with the VLTI/AMBER
     interferometer using the fringe tracker FINITO in the spectral
     domain 1.45-2.50 $\mu$m with a spectral resolution of $\approx35$ and baselines
     ranging from 15 to 88 meters. We perform independent image
     reconstruction for different wavelength bins and fit the
     interferometric data
     with a geometrical toy model. We also compare the data to 1D dynamical models of Miras
     atmosphere and to 3D hydrodynamical simulations of red supergiant (RSG)
     and asymptotic giant branch (AGB) stars.
}
   {Reconstructed images and visibilities show a strong wavelength
     dependence. The H-band images display two bright spots whose positions
     are confirmed by the geometrical toy model. The inhomogeneities are
     qualitatively predicted by 3D simulations. At
     $\approx2.00$ $\mu$m and in the region $2.35-2.50$ $\mu$m, the photosphere
     appears extended and the radius is
     larger than in the H band. In this spectral region, the geometrical toy
     model locates a third bright spot outside the photosphere that can be a feature of the molecular layers.
     The wavelength dependence of the visibility
     can be qualitatively explained by 1D dynamical models of Mira
     atmospheres. The best-fitting photospheric models show a good
     match with the observed visibilities and give a photospheric diameter of $\Theta=8.82\pm0.50$ mas. The H$_2$O molecule seems to be the dominant absorber in the molecular layers.}
   {We show that the atmosphere
 of VX Sgr rather resembles Mira/AGB star model atmospheres
 than RSG model atmsopheres. In particular, we see molecular
 (water) layers that are typical for Mira stars.}


   \maketitle
%

\section{Introduction}

VX Sagittarii (HD 165674) is a cool semi-regular
variable with a long mean period of 732 days \citep{1987IBVS.3058....1K}. \cite{1982MNRAS.198..385L} reported a spectral type varying
from M5.5 (near the time of visual maximum) to M9.8 (at minimum
light).  \cite{1982MNRAS.198..385L} determined
that the effective temperature of VX Sgr is ranging between 3300 and
2400 K (maximum
to minimum light). \cite{2007A&A...462..711G} find $T_{\rm{eff}}$=2900 K at the time of their high-resolution spectroscopic
observation, when the star was near minimum light (AAVSO\footnote{www.aavso.org}).
\cite{1982MNRAS.198..385L} found that VX Sgr exhibits stronger CN and VO bands with respect to Mira variables with similar temperature. Enhanced CN absorption is an indicator of high luminosity in RSGs of earlier type and, together with VO, also of S stars. \cite{2000A&AS..146..437S} categorized  VX Sgr as an oxygen-rich star and
found a strong silicate feature at 10 $\mu$m, that indicates a dusty circumstellar environment. Using aperture-masking and IR/optical-telescope array interferometry at 2.16 $\mu$m, \cite{2004ApJ...605..436M} revealed that VX Sgr exhibits a dusty environment with a flux contribution of about 20$\%$ in the K band and some evidence of departure from circular symmetry, even if they could not place strong limits on possible asymmetries because of calibration uncertainties. The dusty environment is confirmed by HST images \citep{2006AJ....131..603S}. VX Sgr's circumstellar environment is the result of the heavy mass loss experienced by the star \citep[$1.3\cdot10^{-5}$M$_\odot$yr$^{-1}$, CO measurements; ][]{1989ApJ...336..822K}. The mass-loss process appears to be particularly asymmetric for the inner regions \citep{1986MNRAS.220..513C}. Using AAVSO data, \cite{2005PASJ...57..341K} show that the optical light curve has a much
smaller amplitude of about 2 mag in the years 1998-2003, much less than the usual 6-7 mag.
An examination of AAVSO data shows that the decrease to this smaller amplitude has happened several times in the last 70 years,
and that the star is probably currently in that state.

The classification of VX Sgr as a red supergiant or an AGB is thus not firmly established.
A further constraint can be brought by an estimate of its luminosity, in order
to better ascertain its position in the HR diagram. \cite{1972ApJ...172...75H} placed VX Sgr in the 
 vicinity of the Sgr OB1 cluster at 1.7 kpc;
 \cite{2003MNRAS.344....1M} found 1.8$\pm0.5$ kpc measuring the GHz
 H$_2$O maser expansion; \cite{2007ChJAA...7..531C} reported a distance
 of 1.57$\pm0.27$ kpc using 43 GHz SiO maser proper motions; finally,
  the trigonometric parallax of Hipparcos
 \citep{2007A&A...474..653V}
 gives a distance of 0.262(+0.655/-0.109) kpc, probably unreliable due to the size and asymmetry
of the stellar photosphere.

Using AAVSO data, we find VX Sgr was at maximum luminosity during our
observations, and we assume a $T_{\rm{eff}}$ of 3200 to 3400K. With
the 2MASS K magnitude \citep{2003tmc..book.....C}, assuming
  a distance d=1.7 kpc, and using data for Galactic red supergiants
from \cite{2005ApJ...628..973L}, we derive a luminosity  $\log L/L_\odot=5.25\pm0.25$ ($M_{\rm{bol}}=-8.4\pm0.6$). The error bar accounts for uncertainties in the photometry, and in the assumed $T_{\rm{eff}}$ at the time of our observation, impacting the
bolometric correction at K. Circumstellar emission in IR may increase further the
luminosity by a few tenths of magnitude. Putting the star at 1.3 kpc would decrease the estimated luminosity to $\log L/L_\odot=5.00\pm0.25$.
The radius is then about 1200 $R_\odot$, and $A_v$=2 to 4.
This is a too high luminosity for an AGB star \citep[e.g.,][]{1993ApJ...413..641V}. Even compared to so-called super-AGB stars, where the most recent models show a maximum of $\log
L/L_\odot\approx4.8$ with typical masses
ranging between $\approx7-11$ $M_\odot$ \citep{2006A&A...448..717S} and $\log L/L_\odot\approx5$ \citep{2008ApJ...675..614P}, VX Sgr's luminosity is extremely high.
\cite{2009ApJ...705L..31G} found AGB stars with
similar luminosities and masses of $\approx$6-7 $M_\odot$ showing Rb enhancement in the Magellanic Clouds, and they argue that
these AGB stars may be more luminous due to a contribution from Hot Bottom Burning.
However,  \cite{2007A&A...462..711G}, using synthetic spectra based on classical hydrostatic model atmospheres for cool stars with extensive line lists, found VX Sgr to be the only not Li-rich, long-period,
high OH expansion velocity star of their Galactic AGB sample. On the other hand VX Sgr's
low effective temperature, and large V variability are quite untypical for an RSG, although
\cite{2007ApJ...667..202L} found high-variability, low-$T_{\rm{eff}}$ RSGs in the Magellanic Clouds. \cite{1997A&A...327..224H} studied the pulsations properties in red supergiants from 10 to 20 $M_\odot$ with high luminosity to mass ratio and show that very large pulsation periods, amplitudes and mass-loss rates may be expected to occur at and beyond central helium ex-haustion over the time-scale of the last few $10^4$ years. This could lead to an overall dimming of the star after a period of stronger oscillations subsequent to enhanced mass-loss and ejection of a dust shell that screens the stellar radiation.\\
It appears that the evolutionary status of VX Sgr is still not well established, and
more investigations are needed. In particular its chemical composition should be scrutinized.

We discuss here interferometric observations of VX Sgr made with the VLTI/AMBER instrument
in the near IR. The aim of this paper is to study the continuum forming photosphere, the existence, and the characterization of molecular layers of VX Sgr probing different wavelengths in the H and K bands.

\section{Observations and data reduction}

We obtained near-infrared interferometry data of VX Sgr with the Very
Large Telescope Interferometer \cite[VLTI,][]{2008SPIE.7013E..11H}
using the near-infrared beam combiner AMBER
\citep{2007A&A...464....1P} that simultaneously covers the J, H, and K
bands with a spectral resolution of $\approx35$. VX Sgr has been
observed in less than 1.5 months using the AT configurations:
A0-D0-H0, D0-H0-G1 and E0-G0-H0. The fringe tracker FINITO
\citep{2008SPIE.7013E..33L} has been used for all the
observations. In addition to the science target, three calibrator
stars have been observed close in time and interleaved with VX Sgr:
HD169916 (K0IV), HD146545 (K5III) and HD166295 (K2III/IV). The
calibrator diameters were retrieved from \cite{2005A&A...431..773R},
\cite{2002A&A...393..183B} and \cite{2002A&A...386..492R}. The
diameter errors are of the order of $1\%$. Details are reported in
Table~\ref{log} with the projected baseline lengths (B$_p$), the
position angles (PA$_p$), the spectral interval, the calibrators used,
the optical seeing, and the coherence time.

Raw visibilities and closure phases were computed with the latest
version of the \emph{amdlib} package \cite[release
2.2,][]{2007A&A...464...29T} and the \emph{yorick} interface provided
by the Jean-Marie Mariotti Center. Individual frames were averaged after
selecting the $20\%$ best frames based on the fringe SNR only and with
a piston smaller than $\pm20$ $\mu$m. We decided to discard J band
fringes because the data quality was significantly worse than for
longer wavelengths. 

\begin{figure}
   \centering
        \includegraphics[width=1.0\hsize]{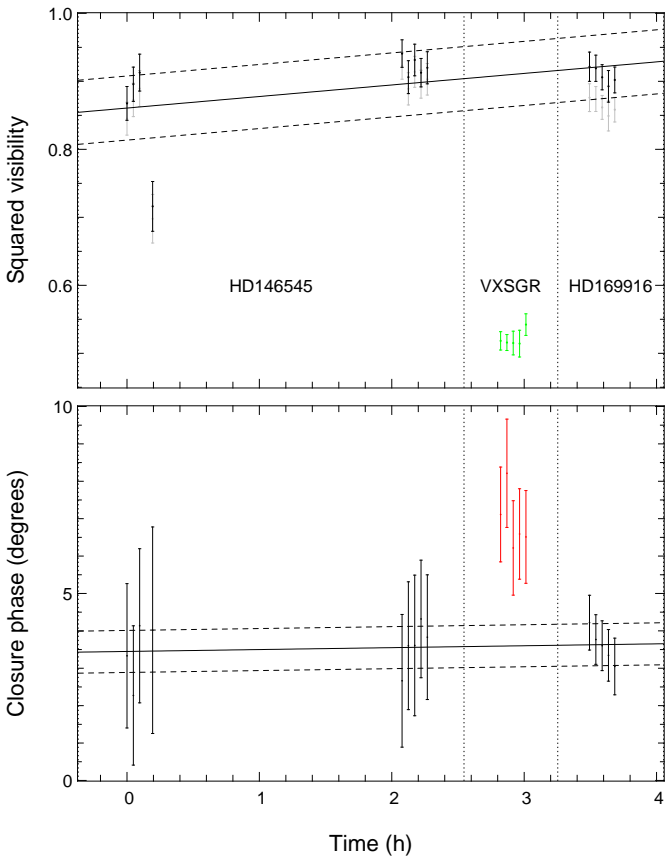}\\
      \caption{Transfer function for the night 2008-07-03. \emph{Top
          panel:} squared visibilities averaged over the region
        2.1-2.2$\mu$m for both the calibration stars (grey with error
        bars) and the science star (green with error bars). The
        computed transfer function is approximated by a linear fit
        (solid black line), and dashed lines display the scatter.
        \emph{Bottom panel:} closure phase averaged over the same
        spectral region for both the science target (red) and
        calibrators (black).
                 }
        \label{tfplot}
   \end{figure}

In addition we used the addons by \citet[][release
1.53]{2008SPIE.7013E.132M} to calibrate the datasets. The transfer
function measurement for one night is shown in Fig.~\ref{tfplot}. Both
science and calibrator targets have the same detector integration time
of 0.05 ms, and after the calibration, VX Sgr data have been averaged.
The error bars on the calibrated visibilities include the statistical
error of averaging the single frames, the errors of the calibration
stars' angular diameters, and the scatter of the transfer function
measurements. This scatter (top panel of Fig.~ \ref{tfplot}) is much
larger than the internal errors, computed by the reduction
software. This basically means that the visibilities errors
  (between 0.05 and 0.1) we use in this paper reflect calibration
  issues affecting simultaneously a whole range of wavelengths, while
  the wavelength-to-wavelength error is much smaller (typically 0.01
  to 0.05): while the visibility errors seem large in a single
  dataset, the wavelength-variation errors of the same dataset are
  small.

The absolute wavelength correction has been done using the telluric
Kitt Peak spectra which we convolved to match the spectral resolution
of the AMBER data. In VX Sgr data, the band gaps (i.e., between J and
H and H and K) are visible and we made a linear two-component
adjustment of the wavelength scale which gave a systematic offset of
-0.21 $\mu$m with respect to the initial AMBER table and a $7\%$
wavelength stretch. 

\begin{figure}
   \centering
        \includegraphics[width=1.0\hsize]{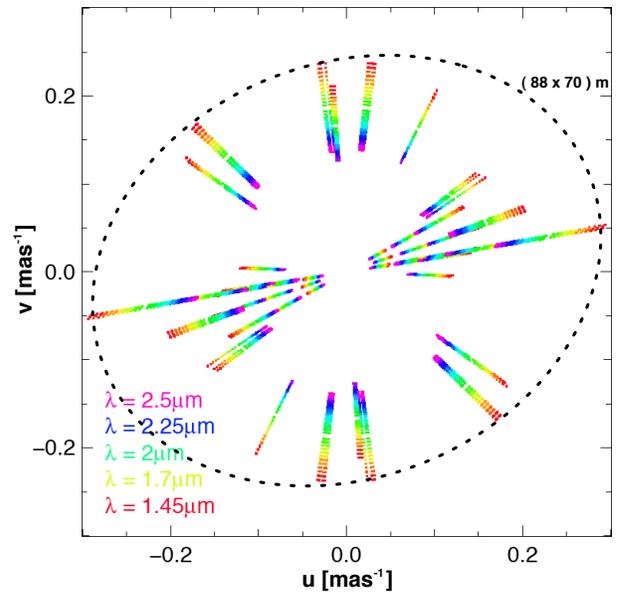}\\
      \caption{UV-coverage for of the AMBER observations of VX Sgr. The radial extension of the uv tracks reflects the spectral
coverage of our AMBER measurements, covering the
H and the K bands.
                 }
        \label{uvplane}
   \end{figure}

Fig.~\ref{uvplane} shows the final UV-plane coverage of all
observations that successfully passed all steps of the data reduction
and calibration quality control. The north-west south-east direction
is not completely covered because of the actual AT geometry, while the
east-west direction is particularly favored.
The star exhibits large wavelength-dependent visibilities and has
a clear non-zero, non-180$^\circ$ closure phase (bottom panel of Fig.~
\ref{tfplot}), evidencing asymmetries in the intensity distribution.

\begin{table*}
\caption{Observations log for the AMBER observations of VX Sgr. All observations were carried out using FINITO and an integration time of 50 ms.}             
\label{log}      
\centering                          
\begin{tabular}{c c c c c c c}        
\hline\hline                 
 {\rm Date} & {\rm B$_{p}$ } & {\rm PA$_{p}$} & Spectral  & Calibrators &Seeing  & Coherence  \\    
  		&	[m]			&			[$^\circ$]		& range [$\mu$m]  &   &    ['']    & time [ms] \\
\hline                        
 	2008-05-24 & H0-D0 (58.76)/D0-A0(29.37)/A0-H0 (88.14) & -99.8/-99.8/-99.8 & 1.8731-2.4766 & HD169916 & 0.65 & 4.5 \\
	2008-05-26 & G1-D0 (71.55)/D0-H0(63.94)/H0-G1 (71.48) & -46.0/70.5/7.0 & 1.4492-2.3398 & HD166295 &0.97 & 6.0 \\
	2008-06-06 & H0-G0 (30.10)/G0-E0(15.06)/E0-H0 (45.17) & -119.0/-119.0/-119.0 & 1.4492-2.4458 & HD169916, & 0.79 & 4.5\\
	    		   & 									        &                                          &                              & HD146545 &   & \\
	2008-06-07 & H0-G0 (28.64)/G0-E0(14.33)/E0-H0 (42.97) & -98.8/-98.8/-98.8 & 1.4655-2.5000 & HD169916 &1.16 & 2.0 \\
	2008-06-08 & H0-G0 (31.98)/G0-E0(16.00)/E0-H0 (47.99) & -108.0/-108.0/-108.0 & 1.4454-2.3708 & HD169916, &0.60 & 2.5\\
	   & 									        &                                          &                              & HD146545 &   & \\
	2008-07-03 & H0-G0 (29.63)/G0-E0(14.82)/E0-H0 (44.46) & -100.0/-100.0/-100.0 & 1.4578-2.2786 & HD169916, &1.29 & 2.0\\
	   & 									        &                                          &                              & HD146545 &   & \\
	2008-07-04& G1-D0 (63.13)/D0-H0(35.39)/H0-G1 (68.78) & -4.3/35.4/68.8 & 1.4454-2.4062 & HD169916 & 0.41 &7.5\\ 
	2008-07-05 &G1-D0 (71.52)/D0-H0(63.92)/H0-G1 (71.46) & -46.1/63.9/6.9 & 1.4511-2.4832 &  HD169916 &0.65 &6.5 \\
	2008-07-06 & G1-D0 (65.26)/D0-H0(55.29)/H0-G1 (71.48) & -55.2/52.6/-7.78 & 1.4540-2.5211& HD169916 &0.60 & 3.5\\
\hline                                   
\end{tabular}
\end{table*}

\section{Image reconstruction}

The first step in our analysis is a chromatic image reconstruction of our data to probe different layers in the photosphere and above.

\begin{figure*}[htbp]
 
   \centering
   	\begin{tabular}{cc}
	\includegraphics[width=0.23\hsize]{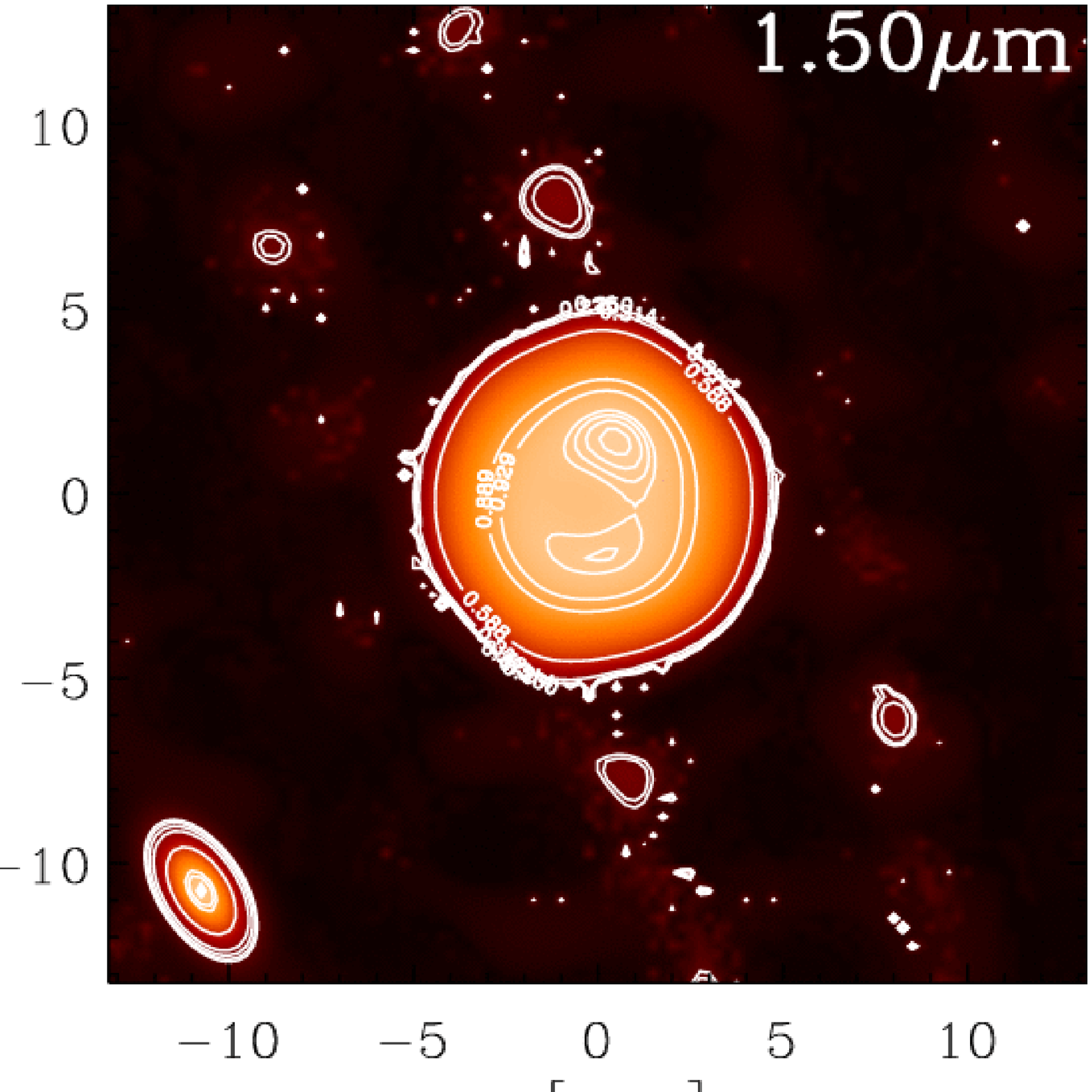}
	\includegraphics[width=0.32\hsize]{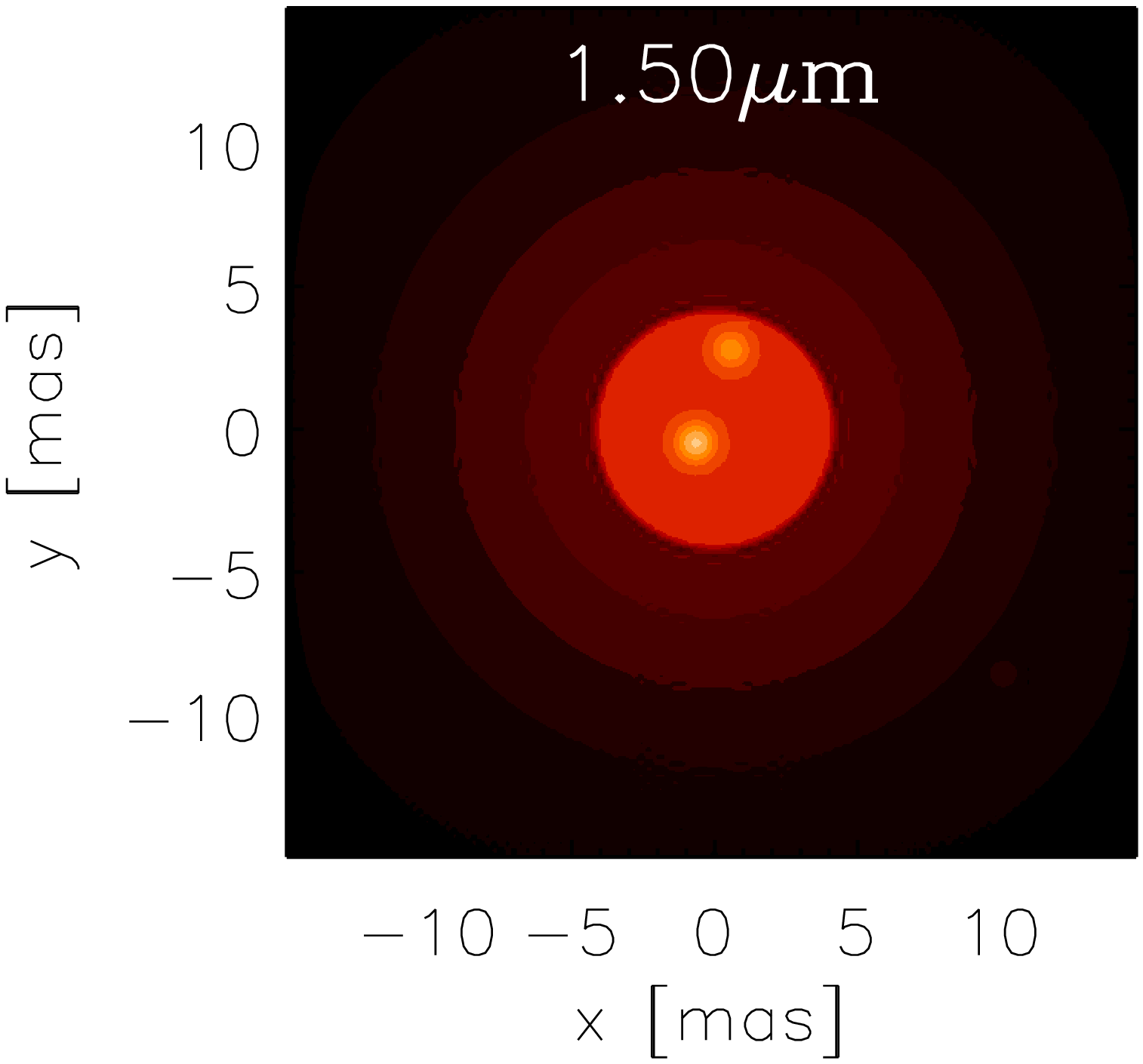}\\
	\includegraphics[width=0.23\hsize]{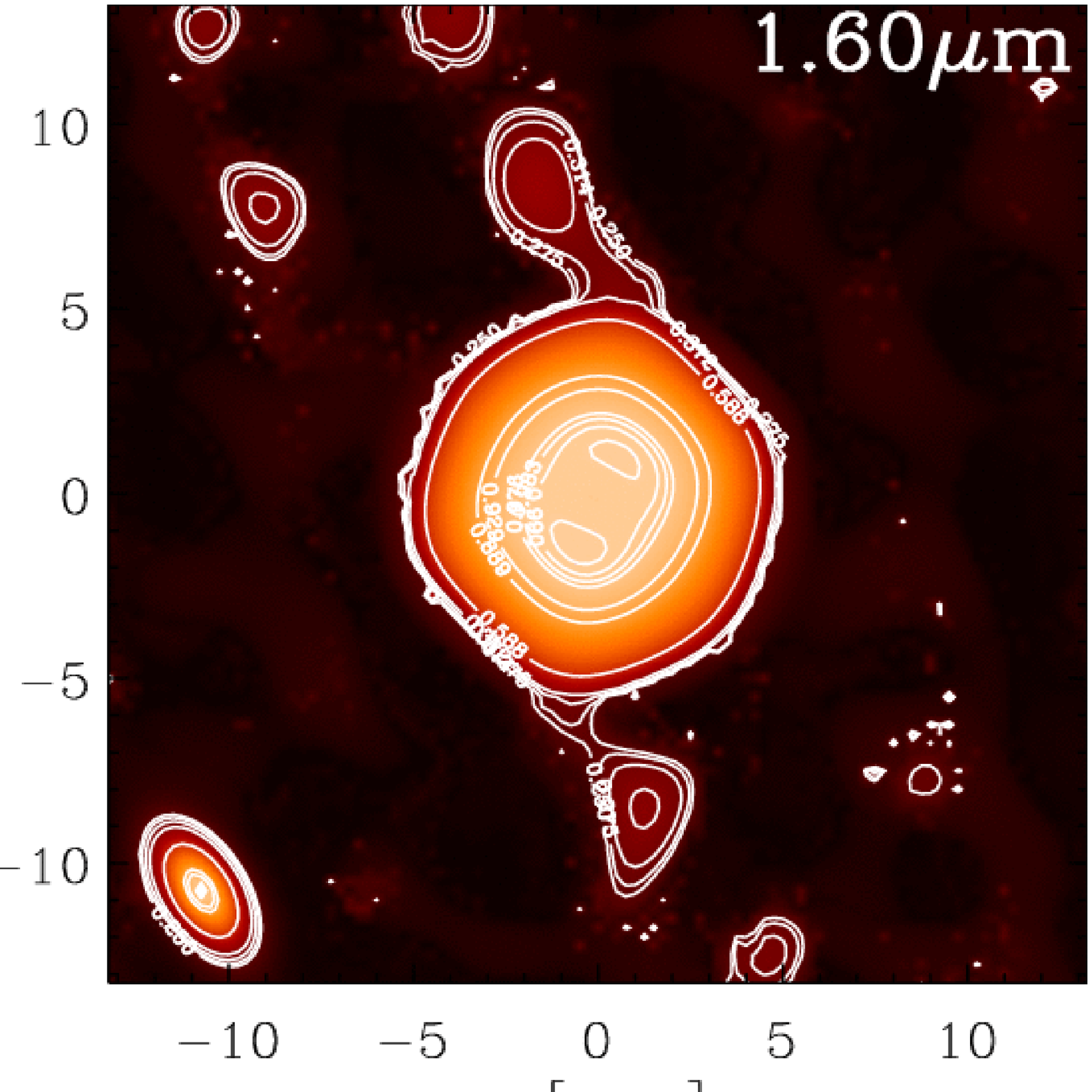}
	\includegraphics[width=0.32\hsize]{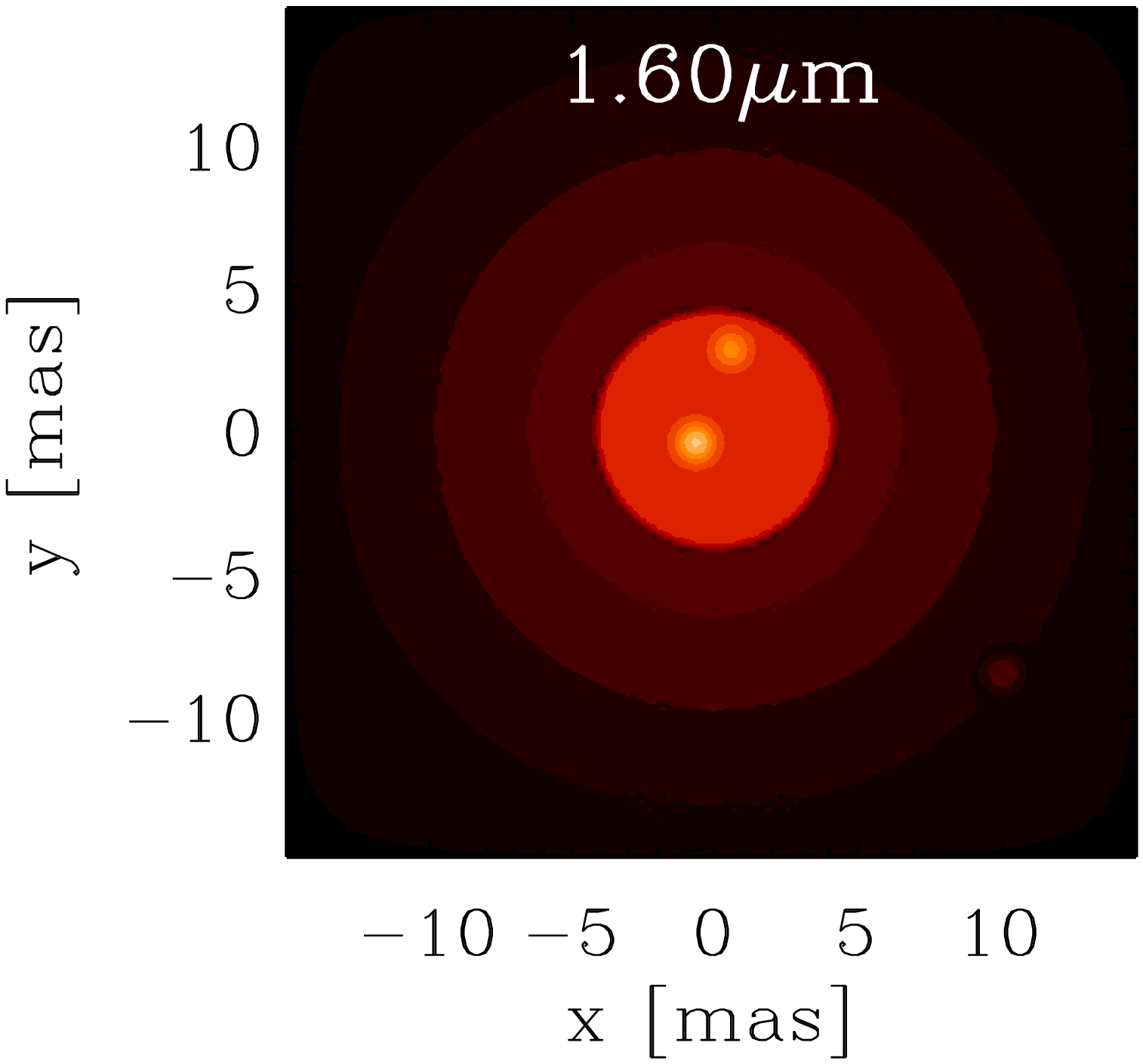}\\
	\includegraphics[width=0.23\hsize]{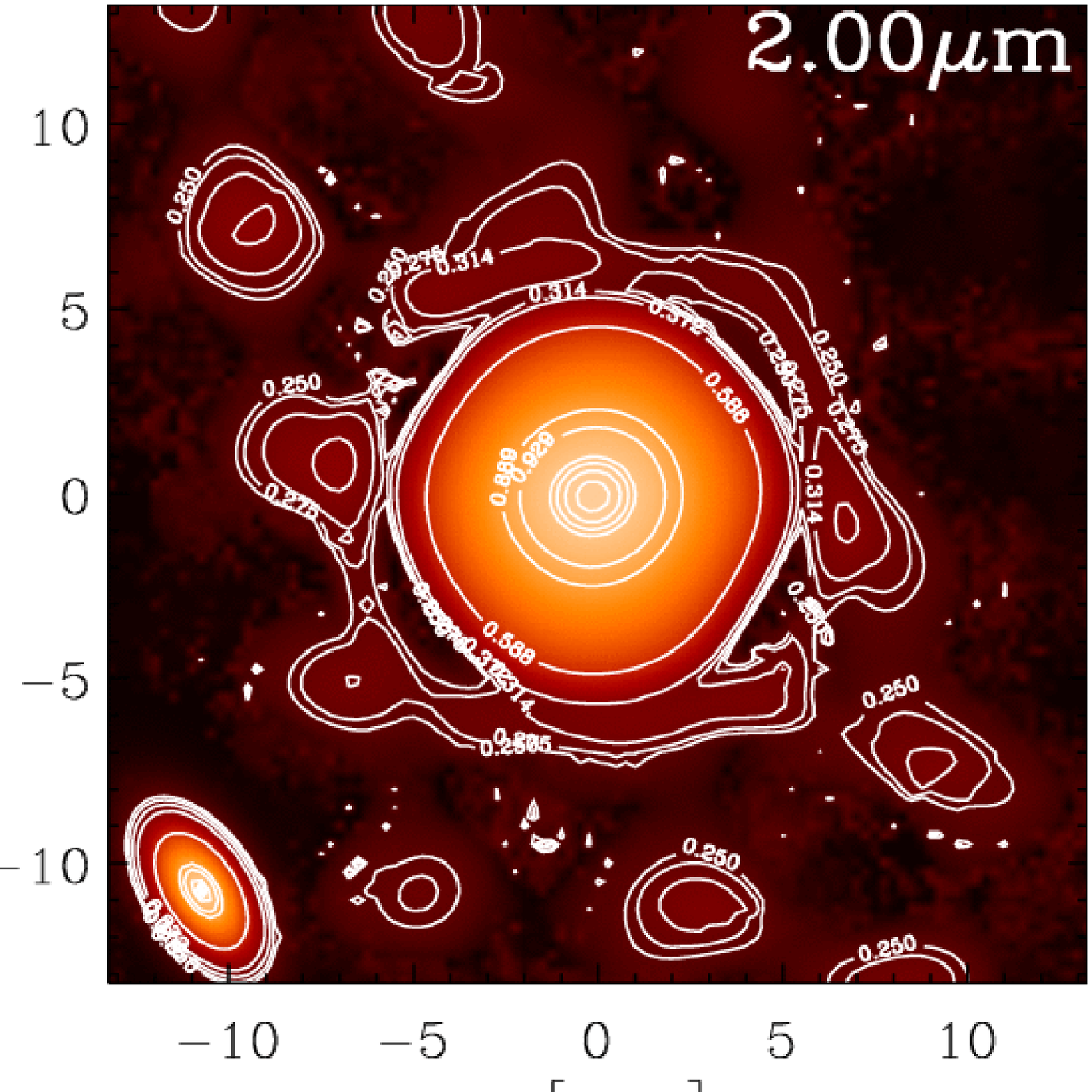}
	\includegraphics[width=0.32\hsize]{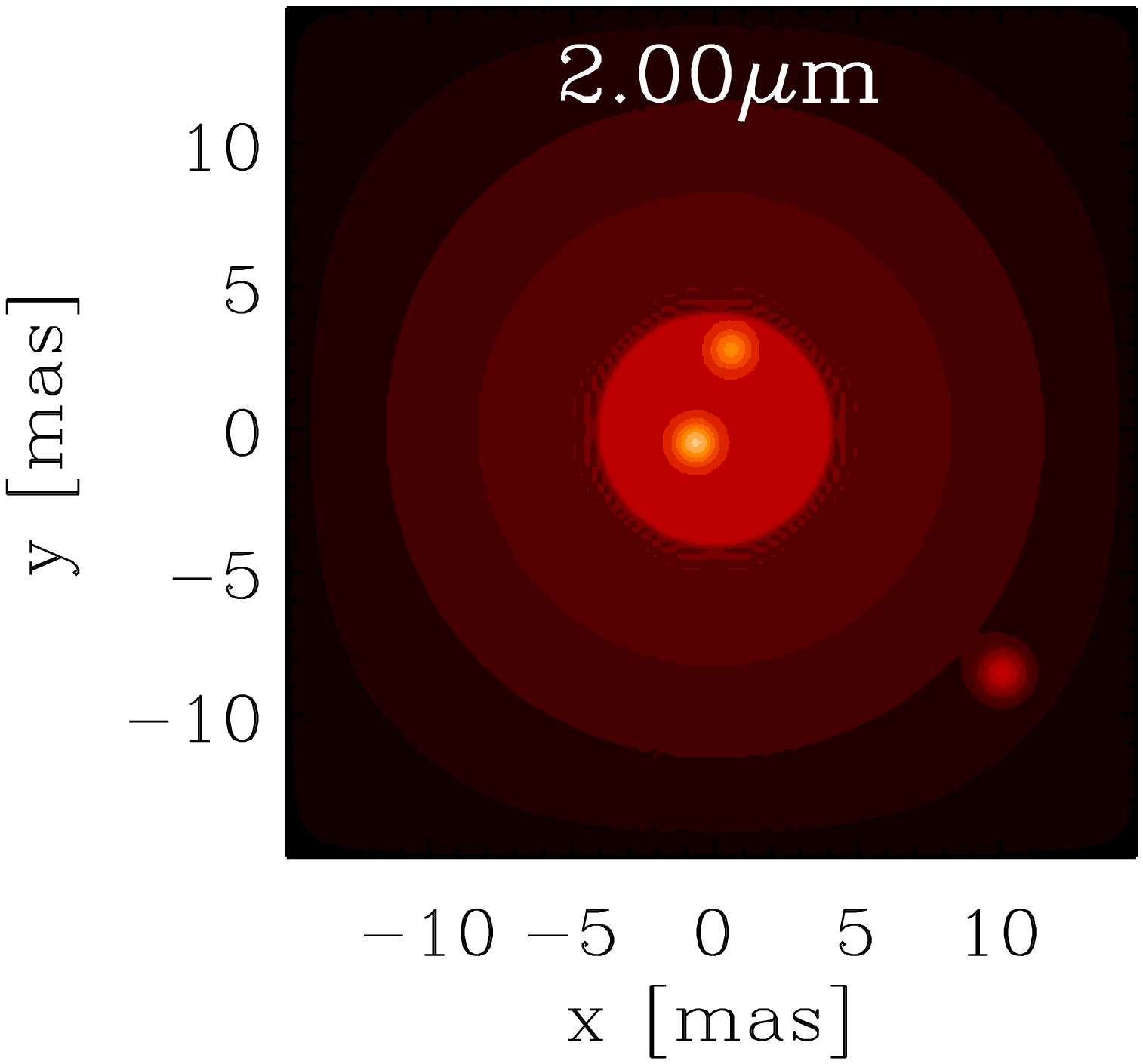}\\
	\includegraphics[width=0.23\hsize]{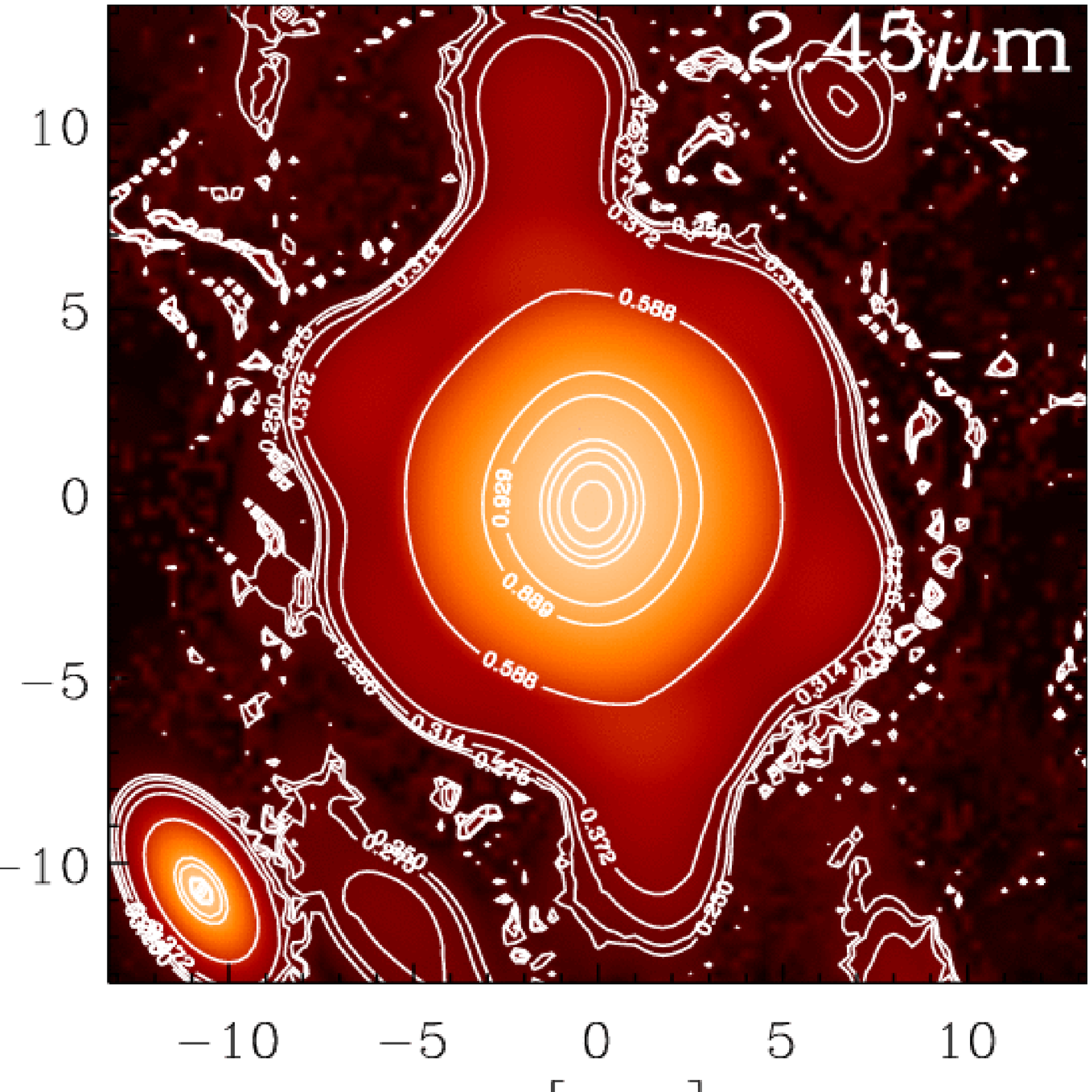}
	\includegraphics[width=0.32\hsize]{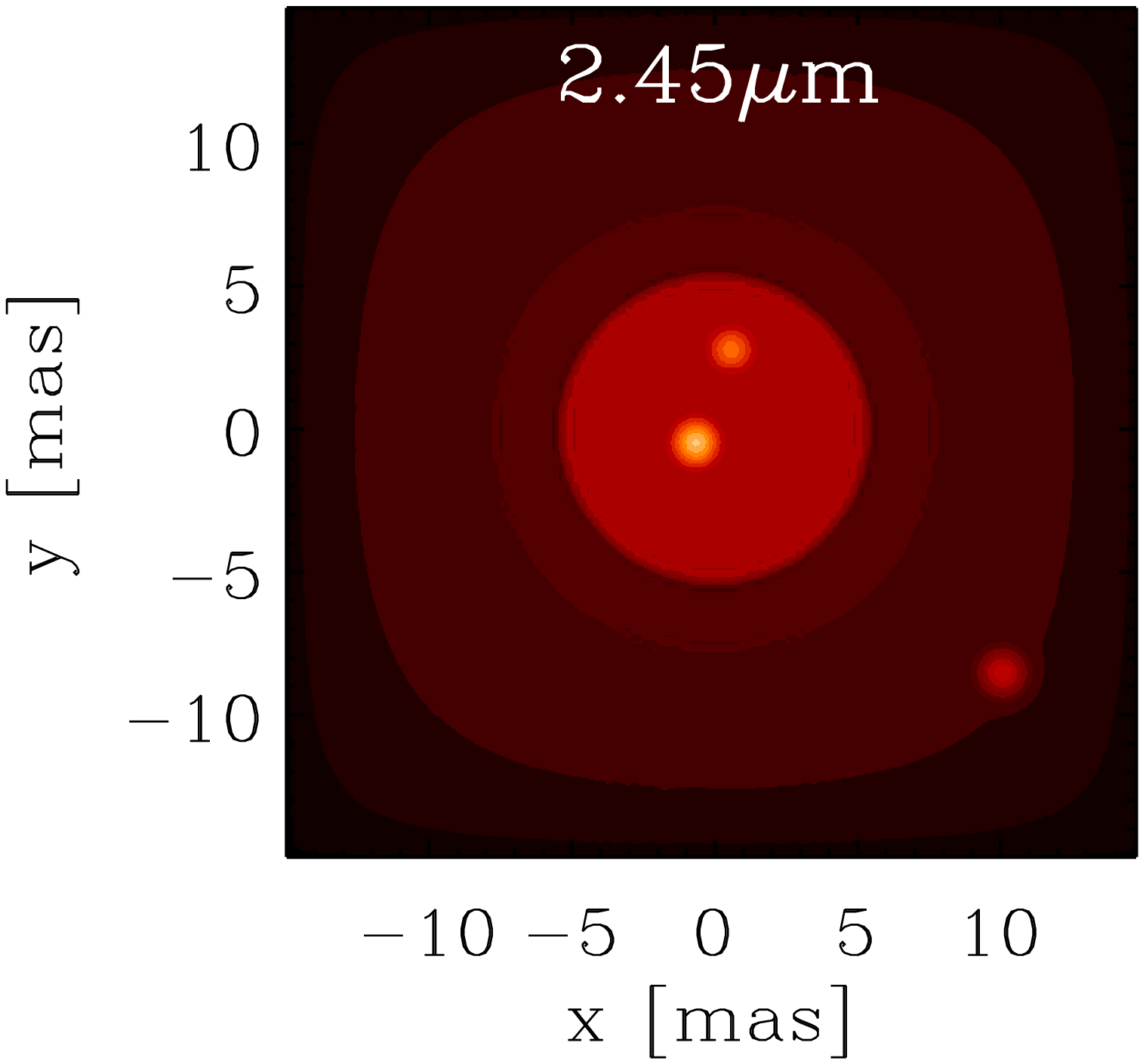}\\
	\includegraphics[width=0.23\hsize]{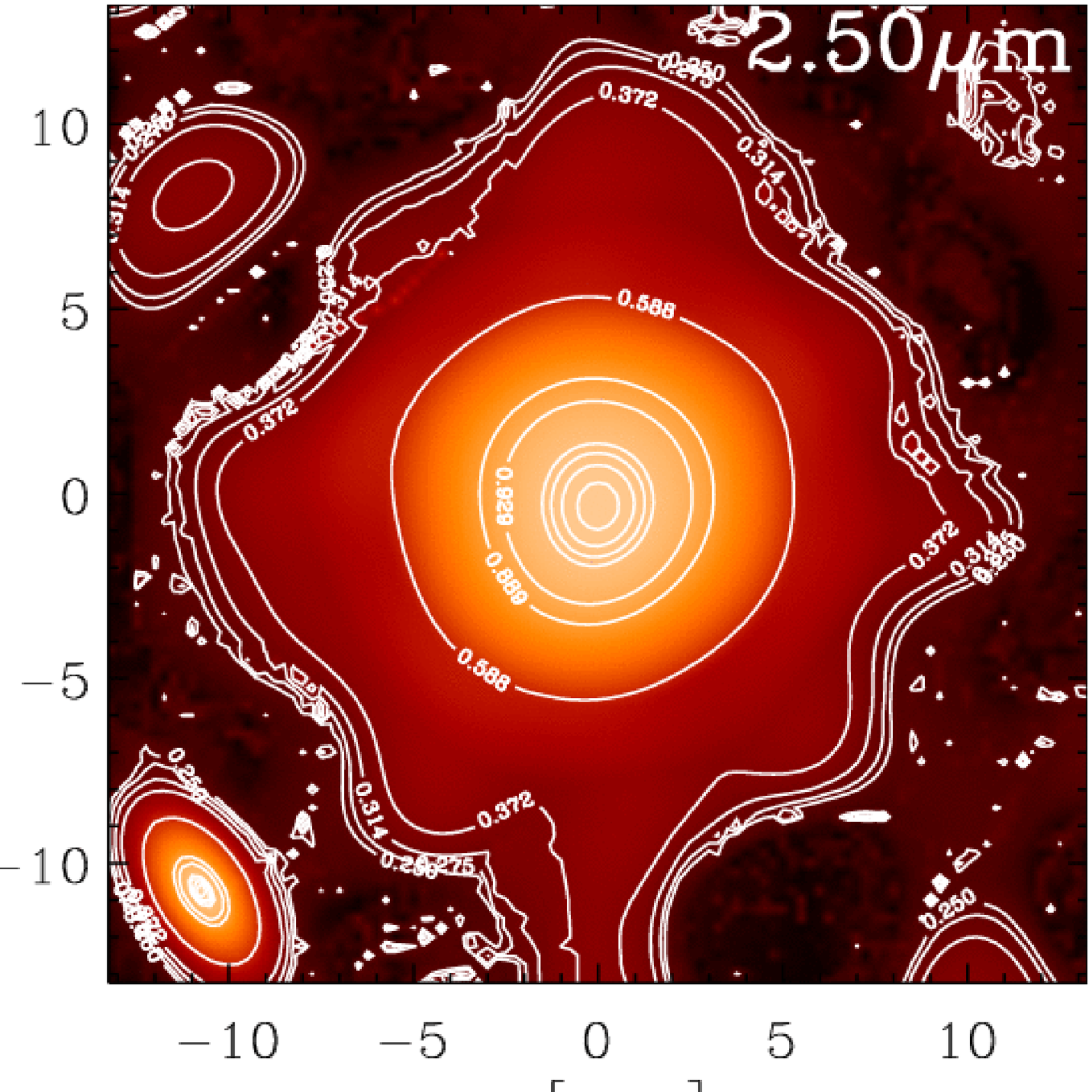}
	\includegraphics[width=0.32\hsize]{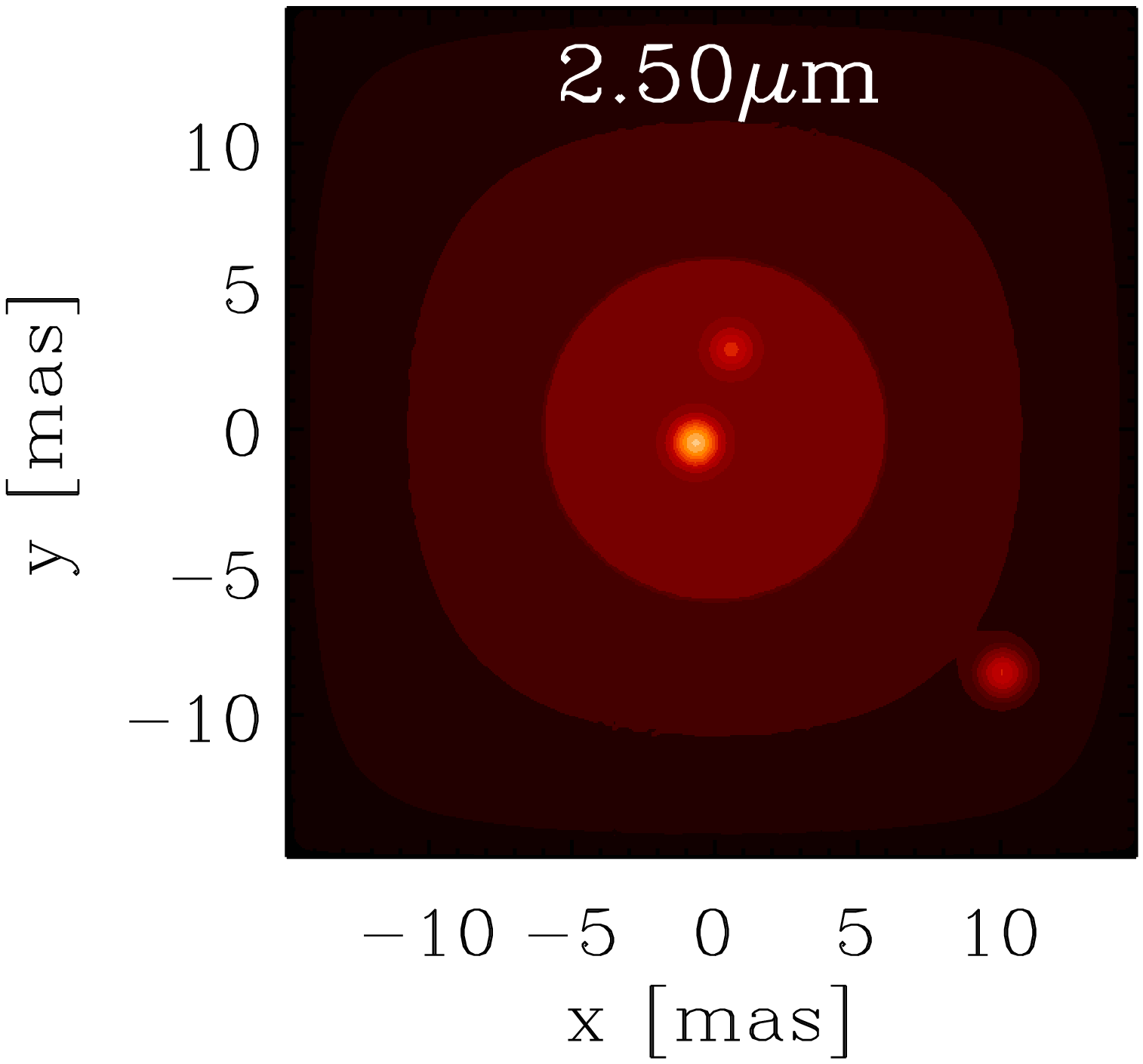}
	\end{tabular}
      \caption{\emph{Left column:} reconstructed images of VX Sgr for several AMBER
        spectral bins across the H and K bands. The intensity, $I$, is
        normalized to the range [0, 1] and plotted as $I^{0.33}$. Some
        contour lines are indicated (0.251, 0.275, 0.314, 0.372, 0.588, 0.889, 0.929,
        0.976,0.983,0.990,0.997). The resolution of the interferometer is illustrated in the bottom
left part of each image by the PSF of a 88$\times$70 m telescope. \emph{Right column}: images representing our best-fitting geometrical model for the same spectral channels. The intensity, $I$, is
        normalized to the range [0, 1] and plotted as $I^{0.33}$.}
        \label{images}

   \end{figure*}

The image reconstruction process is similar to the one performed for T Lep in \cite{2009A&A...496L...1L}. For the reconstruction, we use the MIRA software package \citep{2008SPIE.7013E..43T,2008SPIE.7013E..48C, {2008ISTSP...2..767L}}. The image is sought by minimizing a so-called
\emph{cost function} which is the sum of a regularization term plus
data-related terms. The data terms enforce agreement of the model
image with the measured data (visibilities).
The regularization term is a $\chi^2$ minimization between the
reconstructed image and an \emph{expected} image. The expected image is
issued from a preliminary image reconstruction, strongly constrained by
the assumption of circularity. Each spectral bin has been processed
independently (Fig.~\ref{images}, left column).
The reconstructed images clearly highlight different behaviors across
the wavelength range: in the H band ($\approx1.45-1.80$ $\mu$m), the intensity distribution is inhomogeneous and a bifurcation of the image core 
into a few bright
``spots'' is visible; at $\approx2.00$ $\mu$m and at the upper
K band edge ($\approx2.35-2.50$ $\mu$m) the radius appears extended
and much larger than in the H band.
Artifacts may be introduced by the poor UV-plane coverage in one direction (north-west south-east, see Fig~\ref{uvplane}). However, the detection of inhomogeneities is out of doubt because there are clear signatures in the closure phases (Fig.~\ref{tfplot}).

\section{Geometrical toy models}\label{GEOMOD}

The reconstructed images are not good enough to firmly establish the presence of inhomogeneities on VX Sgr's photosphere because the different sources of errors can smooth out the information.
 Thus, in order to confirm their presence and to constrain their flux relative
to the total flux, we used a model-fitting approach. We approximated the star with a uniform disk and its extended shells
by a Gaussian disk, and we added a series of point sources to model
the spots. We performed a fit using all the observed wavelengths together, but
used five regularly spaced reference wavelengths to effectively
compute the model (1.54, 1.78, 2.03, 2.28, and 2.52 $\mu$m). To match
the data with the model, we interpolated the parameters for other
wavelengths using cubic spline interpolation. The global optimum of the fit was
found using a set of simulated annealing algorithms complemented by
standard gradient descent algorithms \citep[see][for a first
application of this approach]{Millour09}. A model with three spots (i.e., performing a fit with a
uniform disk, a Gaussian disk, and three spots) significantly enhance the fit
($\chi^2\approx25$), especially for the visibilities and closure phases up to spatial frequency of
$\approx240$ cycles/arcsec. Spatial frequencies above this value
correspond to a single measurement at the largest baseline length. 
The details about the fits are reported in Table~\ref{fitlog}.


Our toy model is probably not perfectly describing the
object at the highest angular resolution available, but gives a
resonable fit to intermediate angular resolutions. We also note that two of the bright spots are located at the position of the stellar disk,
and one has to be \emph{outside} the stellar disk
in order to fit the observed closure
phases. We estimate that the upper value of the spots'
flux is $\leq10\%$ of the total flux and that the flux of the spot located
outside the photosphere has very likely a flux equal to zero in the H
band. Fig.~\ref{images} (right column) displays the appearance of the
toy model at some selected wavelengths. As a final remark, these images resemble the reconstructed images in the left column of the figure. We also note that the bright spot located outside the photosphere is particularly visible in the lower right part of each panel of the reconstructed images at lower wavelength and somehow less evident at longer wavelength even if it is expected by the geometrical model.

\begin{table*}
\begin{minipage}[t]{\columnwidth}
\caption{Parameters of the best geometrical toy model ($\chi^2\approx25$).}             
\label{fitlog}      
\centering                          
\renewcommand{\footnoterule}{} 
\begin{tabular}{|c|cc|cc|ccc|ccc|ccc|cccc|}        
\hline\hline                 
 $\lambda$ & \multicolumn{2}{c}{Gaussian disk} & \multicolumn{2}{c}{Uniform disk} & \multicolumn{3}{c}{Point source} & \multicolumn{3}{c}{Point source} & \multicolumn{3}{c|}{Point source} \\
               & size  & flux             & size &
    flux & x  & y  & flux &  x  & y  &
    flux & x  & y  & flux \\ 
 $[{\mu}\mathrm{m}]$ & [mas] & [\%] & [mas] & [\%] & [mas] & [mas] & [\%] & [mas] &
     [mas] & [\%] & [mas] & [mas] & [\%] \\
\hline
     1.54 & $\leq20$     & 26.7 & $8.3\pm2.5$  & 67.0  & $0.6\pm0.8$ & $2.8\pm0.5$ & $\leq$10   & $-0.7\pm0.7$ & $-0.5\pm0.5$ & $\leq$10    & $10\pm4$ & $-8.5\pm3.5$ & 0             \\
     1.78 & $\leq20$     & 32.1 & $8.1\pm1.5$  & 59.3  & -    & -    & $\leq$10   & -    & -     & $\leq$10    & -    & -     & 0              \\
     2.03 & $\leq20$     & 47.3 & $8.0\pm0.6$  & 43.8  & -    & -    & $\leq$10   & -    & -     & $\leq$10    & -    & -     & $\leq$5       \\
     2.28 & $\leq30$     & 46.1 & $7.7\pm0.4$  & 47.5  & -    & -    & $\leq$10   & -    & -     & $\leq$10    & -    & -     & $\leq$10            \\
     2.52 & $28.7\pm26$  & 66.9 & $12.4\pm1.9$ & 22.9  & -    & -    & $\leq$10   & -    & -     & $\leq$10    & -    & -     & $\leq$10            \\
\hline                        
 
\hline                                   
\end{tabular}
\end{minipage}
\end{table*}

What we can conclude from this model-fitting approach is the
following:
\begin{itemize}
\item a model with three spots can explain the visibilities
and closure phases for spatial frequencies $\leq240$
cycles/arcsec. Higher spatial frequencies are probed in our dataset
by only one observation and indicate a more complicated object shape
at smaller scales;
\item The best-fitting model locates two of the spots inside the supposed
photosphere of the star, but the third spot, mandatory to explain the
closure phase deviations, is located outside the photosphere and it
is brighter at longer wavelenths;
\item The spots have individual fluxes which, at all wavelengths, do not exceed 10\% of the
total flux of the star, for all wavelengths.
\end{itemize}

\section{Model fitting}\label{fittingSect}

\begin{figure}
   \centering
        \includegraphics[width=1.0\hsize]{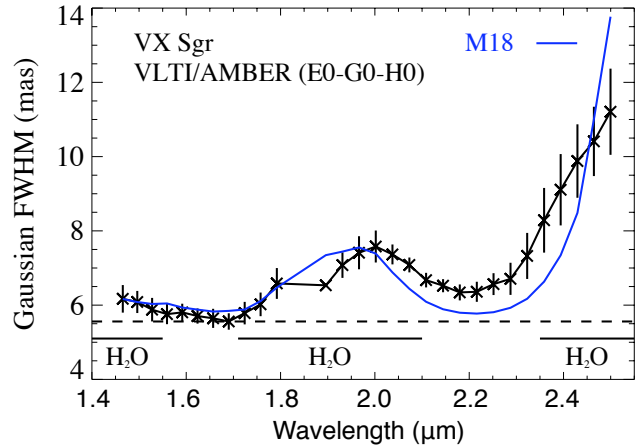}
      \caption{VX Sgr Gaussian FWHM values as a function of the wavelength compared to the model M18 predictions. The dashed line indicates the smallest values. The position of the H$_2$O bands are also indicated \citep[after][]{2000A&AS..146..217L}.
                 }
        \label{gauss}
   \end{figure}

The visibility data of VX Sgr show significant wavelength-dependent features. We fitted the visibility curves and the closure phases using a Gaussian disk for each spectral channel, which gives more robust results than a uniform disk model. Figure~\ref{gauss} illustrates the change of apparent size with wavelength. This is consistent with what is visible in the reconstructed images (Fig.~\ref{images}): the radius gets larger between 1.8 and 2.1 $\mu$m, reaching local maximum at 2.0 $\mu$m, then gets smaller again to a minimum around 2.15 to 2.25 $\mu$m, eventually it becomes much larger toward 2.5 $\mu$m. However, Fig.~\ref{gauss} gives only a rough idea of the apparent diameter of VX Sgr because the intensity profile is not a Gaussian disk.

\begin{figure*}
\hspace*{-4mm}
   \centering
   \begin{tabular}{ccc}
 	\includegraphics[width=0.45\hsize]{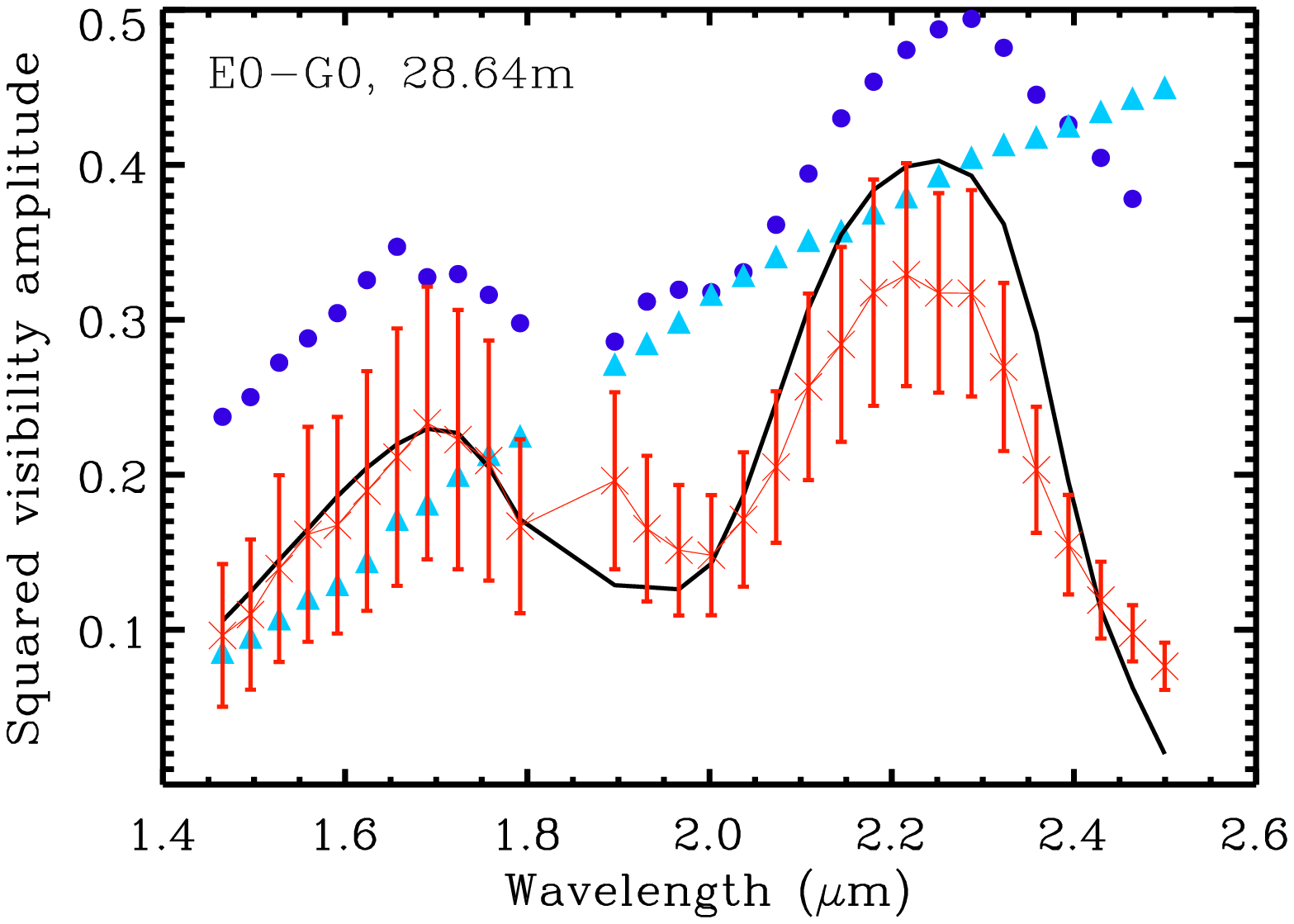}
	\includegraphics[width=0.45\hsize]{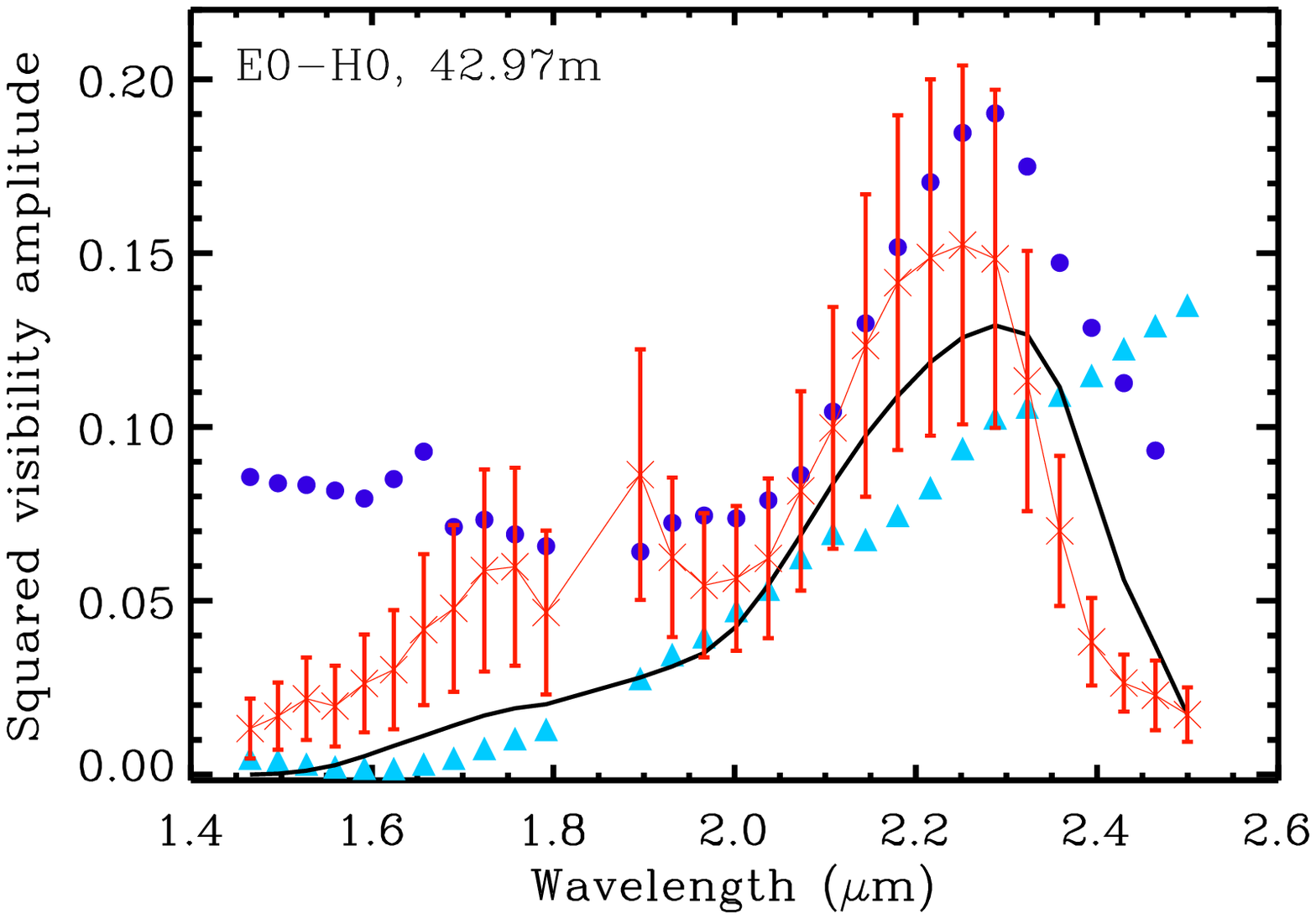}\\
        \includegraphics[width=0.45\hsize]{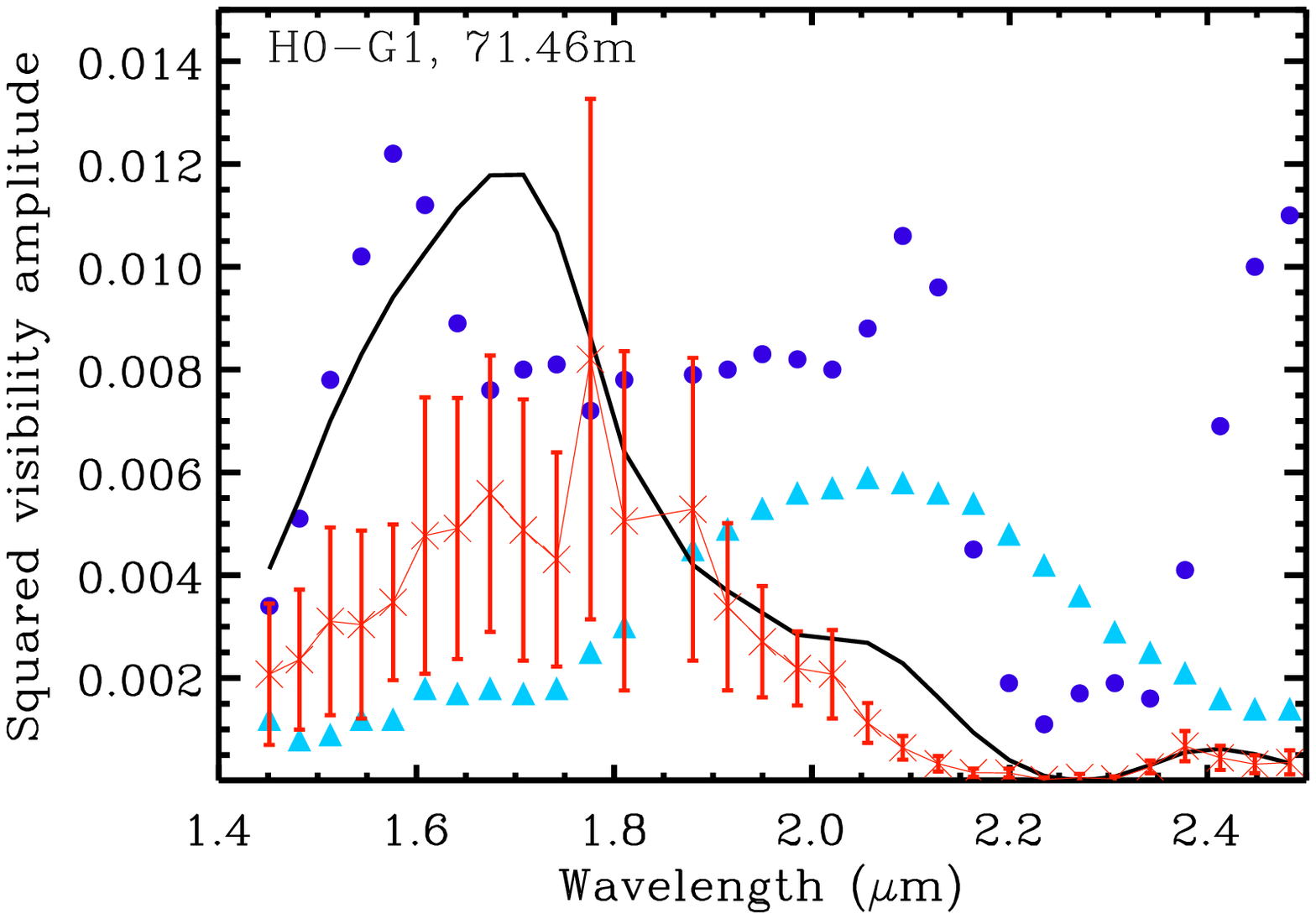}
        \end{tabular}
       \caption{Measured VX Sgr visibility data (red crosses with
         error bars) compared to the visibilities derived from: (i) the 1D Mira atmosphere model
         M18 (model with a phase $\Phi=0.75$),
         black solid line; (ii) the best-matching synthetic image of the
         3D simulation of a RSG star
         (top central and right panel in Fig.~\ref{images2}), light
         blue triangles; (iii) best match synthetic images of the 3D
         simulation of an AGB star (bottom row in
         Fig.~\ref{images2}); blue dots. \emph{Left} and \emph{central} panels belong to the night 2008-06-07, and the \emph{right} panel is for the night  2008-07-03. 
                 }
        \label{visib}
        \hspace*{-4mm}
   \end{figure*}

To reproduce the wavelength dependence, we used the one-dimensional dynamic models of \cite{2004MNRAS.355..444I,2004MNRAS.352..318I} for oxygen-rich Mira stars that include the effect of molecular layers in the outer atmosphere. These are self-excited dynamic models whose grey atmospheric temperature stratification was
re-computed on the basis of non-grey extinction coefficients that contain all
relevant molecular absorbers (e.g., H$_2$O, CO, TiO; solar abundances)
but do not contain dust. The stellar parameters were 
assumed to be close to those of the M-type Mira prototypes o Cet and R Leo. 
This model series has been successfully used by
\cite{2007A&A...470..191W,2008A&A...479L..21W} to explain AMBER
observation of the Mira star S Ori. Due to the poorly understanding of
the VX Sgr stellar parameters, there is no fully consistent model of
this star, and the models we used can show only typical
characteristics of such a star. Figure~\ref{visib} shows the comparison for short (left and central panels) and long (right panel) projected baselines. The wavelength dependence of the
visibility is similar for all the nights. 

For the fit, we followed the fitting procedure used by
\cite{2008A&A...479L..21W} and used only the short baseline data obtained with
  E0-G0-H0, because we do not expect to match details of the intensity
  profile probed at high spatial frequencies with the
  M series that has not been designed to match a star like VX Sgr . The dynamic models predict large changes
in the monochromatic radius caused by the molecular layers above the
continuum-forming region in the stellar photosphere. The bandpasses
around 1.9 and 2.5 $\mu$m are significantly affected by H$_2$O
molecules with some contribution of CO in the H band and in
the long-wavelength part K band. The molecular-band effects change 
significantly with phase as shown in \cite{2008A&A...479L..21W}. The
model M18 (see Table~\ref{modeltable}) provides the best fit to the VX Sgr data among the 20
available phase and cycle combinations of the M series.  

\begin{table}
\begin{minipage}[t]{\columnwidth}
\caption{Parameters of the simulations. All the models have solar
metallicity.}             
\label{modeltable}      
\centering                          
\renewcommand{\footnoterule}{} 
\begin{tabular}{c c c c c c c c}        
\hline\hline                 
 Model & Stellar  & $M$         & $L$      &$R$          & $T_{\rm{eff}}$ & log$g$   \\
name   & type            & $[M_\odot]$ &$[L_\odot]$  & $[R_\odot]$   & [K]        & [cgs] \\
\hline
M18\footnote{1D model; the radius is the Rosseland radius R=0.81R$_p$ at Mira phase $\approx$0.75; \citep{2004MNRAS.355..444I}}    & Mira     & 1.2  & 4840 & 210 & 3310 & - \\
st28gm06n06\footnote{3D model; \cite{2008A&A...483..571F}} & AGB & 1  & 6\,935& 429 & 2542 & -0.83 \\
st35gm03n07\footnote{3D model; \cite{2009A&A...506.1351C}} & RSG & 12 & 93\,000& 832 & 3490 & -0.34 \\
\hline                        
 
\hline                                   
\end{tabular}
\end{minipage}
\end{table}


The M model qualitatively explains the overall wavelength dependence of the
apparent stellar radius (Fig.~\ref{visib},
solid black line). However, while the fit is very good at short
baselines, the comparison is less good at long baselines. Two reasons
can explain this discrepancy: (i) the stellar parameters of the M
models are not well suited for VX Sgr, (ii) there are some
surface inhomogeneities detected in the data, that are not in the
M model. 
 The angular photospheric diameter corresponding to the reference
   radius at 1.04 $\mu$m is
 $\Theta_{\rm{M18}}=8.82\pm0.43$ mas for short baselines where the fit
 is better. The error on the diameter includes systematic calibration
 and model uncertainties. This result is in agreement to what has been found by
 \cite{2004ApJ...605..436M}, $\Theta_{2.16\mu\rm{m}}=8.7\pm0.4$ mas. However, our models do not include dust while \citeauthor{2004ApJ...605..436M} found a dusty environment around the star with a flux contribution of about 20$\%$ in the
K band.
 
We also fitted Gaussian FWHM values to the synthetic visibility values based on the M18 model
intensity profiles for each spectral channel, using the same procedure
that was used to fit the measured data. The comparison between
the fit to the M18 model intensity profiles and the fit to the
measured data is shown in Fig.~\ref{gauss}.

The available Mira models show good agreement with VX Sgr's
data. However, this comparison can just
give a basic qualitative picture of VX Sgr's surface and more detailed
interpretations of these data must be addressed to next generation
models with stellar parameters appropriate for VX Sgr, which are currently not available.

\section{Complementary comparison with three-dimensional simulations}

In this section, we quantify if the observed asymmetries are
consistent with three-dimensional simulations of surface convection in RSG and AGB stars, and if their chaotic photospheric structure is adequate to explain the different surface layers structures observed. The simulation are carried out with CO$^5$BOLD
\citep{2002AN....323..213F, 2008A&A...483..571F}. Parameters of the models are given in
Table~\ref{modeltable}. The pulsations are not artificially added to the models (e.g. by a piston)
but are self-excited. Excitation by not-stationary sonic convective motions are responsible for the pulsations. Molecular opacities are taken into account, but radiation transport is treated in grey approximation, ignoring radiation pressure and dust opacities. Dynamical pressure lets the density drop much slower than expected
for a hydrostatic atmosphere. In our 3D simulations, the average density drops exponentially, and there is no sign of a wind or an extended shell with relatively large
densities. Some more technical information can be
found in \cite{2008A&A...483..571F}, the CO$^5$BOLD Online User
Manual\footnote{www.astro.uu.se/\textasciitilde bf/co5bold$\_$main.html}, and in a forthcoming paper by Freytag (2010, in prep.).

We computed $\approx3500$ synthetic images from these simulations at the
same wavelengths of the observations using our 3D
radiative transfer code OPTIM3D \citep{2009A&A...506.1351C}. Then, we generated
visibility curves using the method described by \citeauthor{2009A&A...506.1351C}. In
the case of the RSG simulation, 
the synthetic images do not change strongly in diameter, shape and
asymmetries across the wavelength range of the observations, and, hence,
the observed wavelength dependence of the visibility is not explained
by this approach (blue triangles in Fig.~\ref{visib}). The synthetic images
of the AGB simulation show a noticeable wavelength dependence and
their agreement with the observed visibilities is slightly better than
the RSG model,
 in particular at short baselines (left and central panel, blue dots
 of Fig.~\ref{visib}). Nevertheless, the two 3D simulations analyzed
 do not provide a better fit to our AMBER data than the Mira models described in
 Sect.~\ref{fittingSect}.

\begin{figure}
   \centering
   	\begin{tabular}{ccc}
 \includegraphics[width=0.7\hsize]{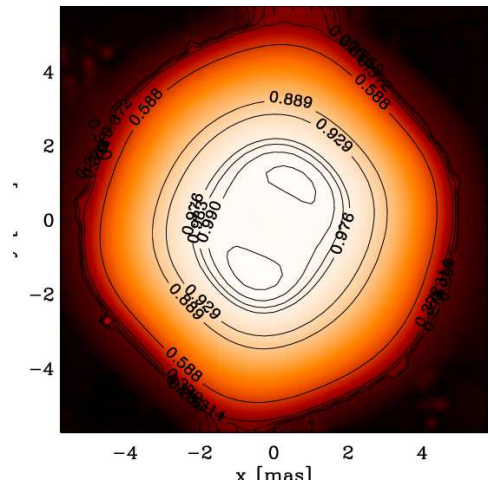}
	\end{tabular}
      \caption{Reconstructed image at 1.6 $\mu$m with the intensity normalized to the range [0,
          1] and plotted as $I^{0.33}$. The contour lines are the same as in Fig.~\ref{images}.
                 }
        \label{images2}
   \end{figure}

\begin{figure*}
\hspace*{-4mm}
   \centering
   	\begin{tabular}{cc}
    \includegraphics[width=0.35\hsize]{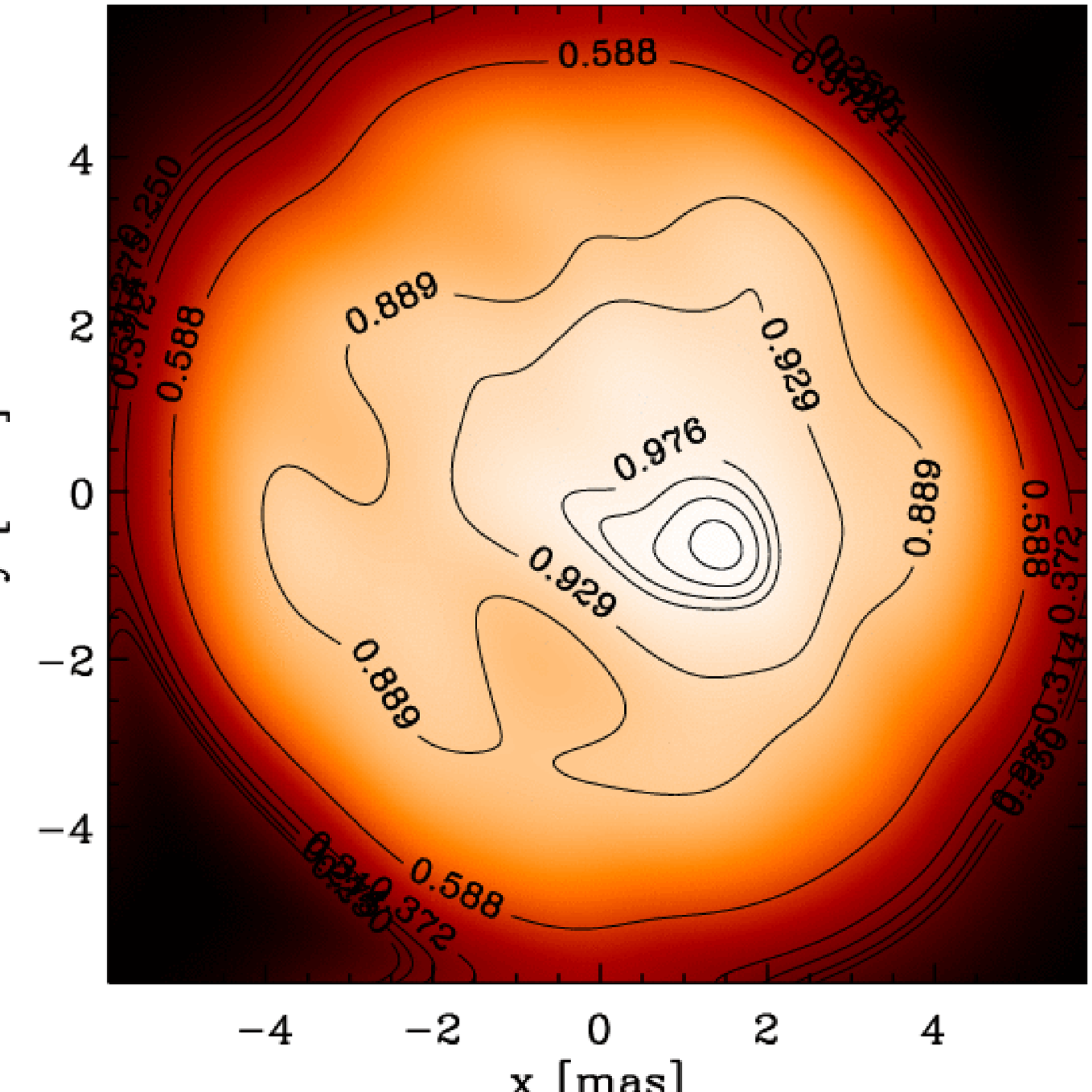}
 \includegraphics[width=0.35\hsize]{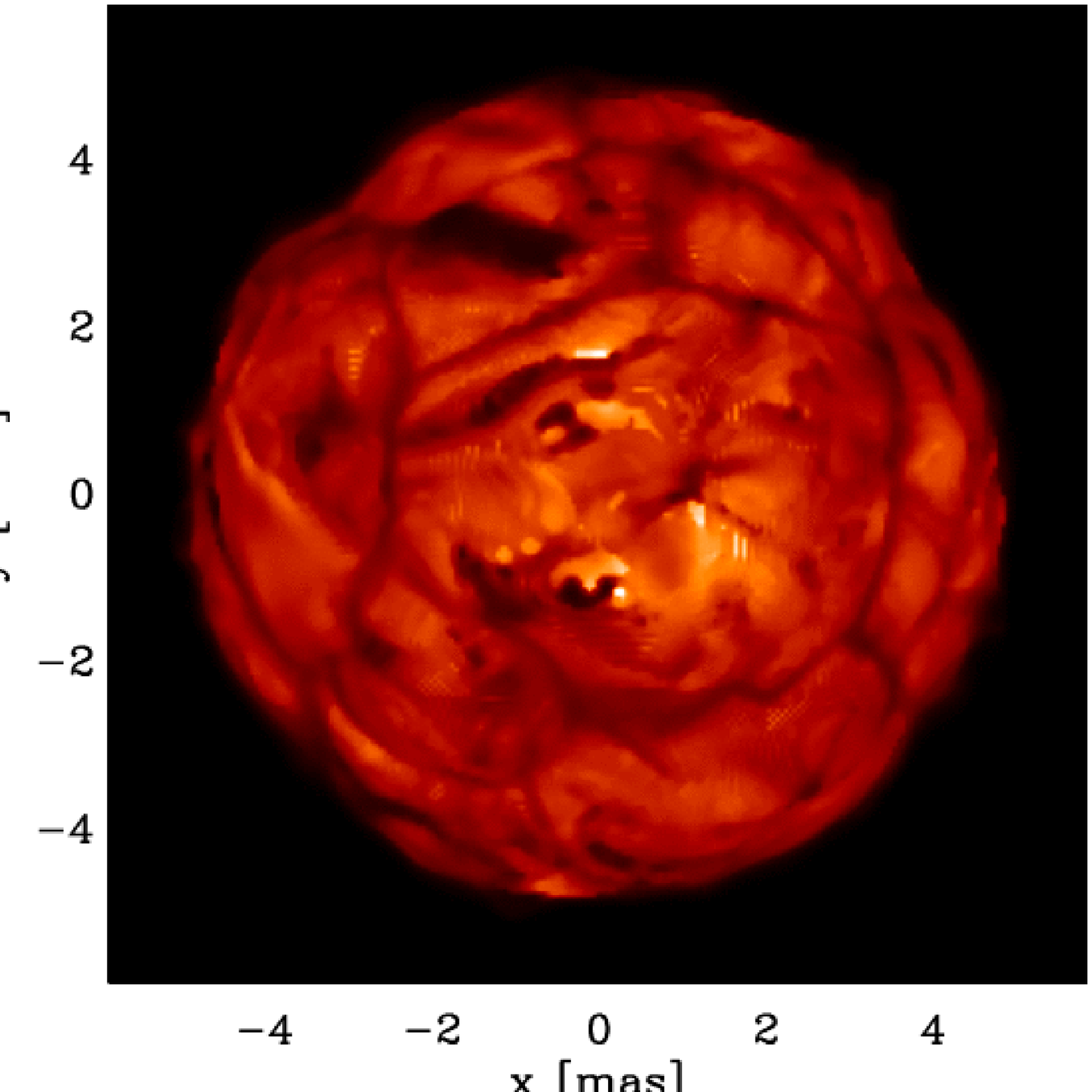}\\
 \includegraphics[width=0.4\hsize]{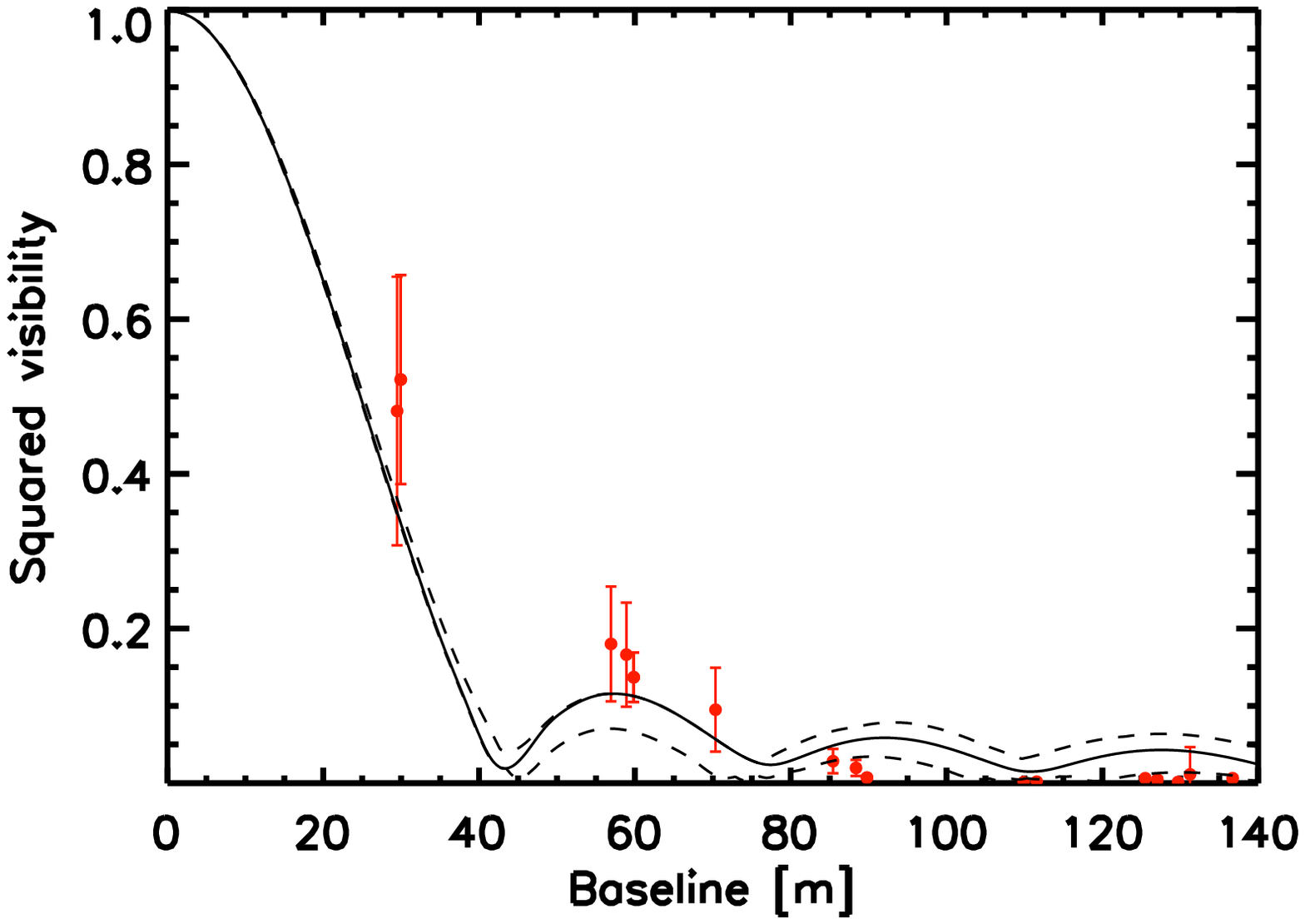}
 \includegraphics[width=0.4\hsize]{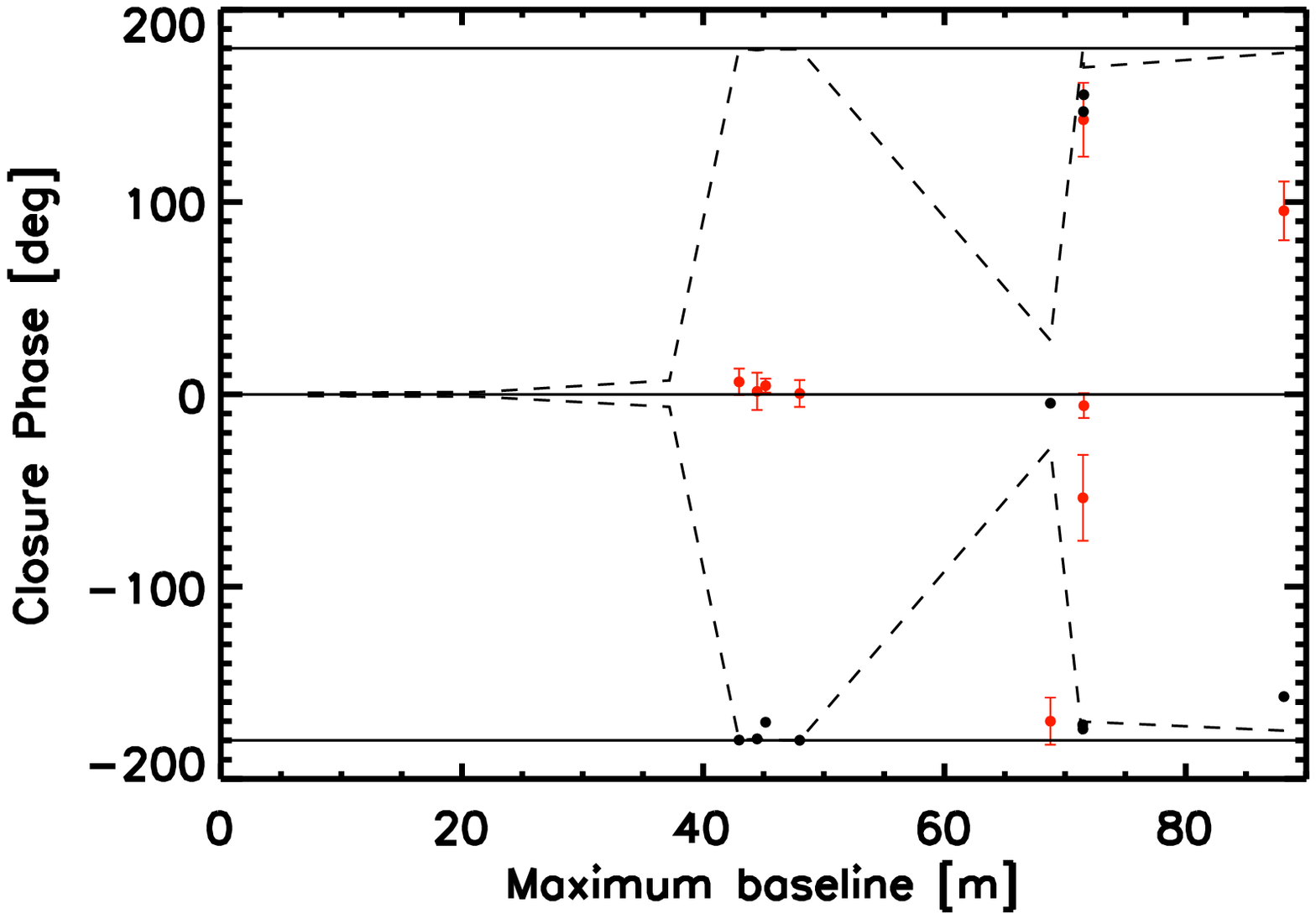}
	\end{tabular}
      \caption{\emph{Top left panel:} synthetic image
        from the RSG simulation at 1.6 $\mu$m, convolved with a 5.9$\times$4.6 mas
        PSF. The intensity is plotted as in Fig.~\ref{images2}. The
        simulation has been scaled in size to match approximately the
        visibility data points. \emph{Top right
          panel: }original RSG simulated image, the range is [0;
          $300\,
          000$]\,erg\,cm$^{-2}$\,s$^{-1}$\,{\AA}$^{-1}$. \emph{Bottom
           panels: } synthetic visibilities and closure phases derived
        from the above image (black solid line and dots) compared to the
        observations (red with error bars) at
        1.6 $\mu$m. The dashed lines indicate the minimum and maximum amplitude of variations of the visibilities and closure phases issued from different rotations of the image. In bottom right panel, the axisymmetric case is
represented by the solid lines.
                 }
        \label{images3}
   \end{figure*}

At last, in order to explain the detected spots on the surface of the
reconstructed images (Fig.~\ref{images}), we compare the
synthetic images corresponding to the best-fitting visibility curves of the 3D
simulations (see Fig.~\ref{visib}) to the
reconstructed image at 1.6
$\mu$m (Fig~\ref{images2}). This wavelength corresponds roughly to the H$^{-1}$ continuous
opacity minimum (i.e., the photosphere becomes more transparent).
The reconstructed image shows two spots on the
surface. Figure~\ref{images3} displays the synthetic images of the RSG star which show
 large convective cells ($\approx4-5$ mas), and on top of that there are
 small-scale granules ($\approx0.5-1$ mas). The observed visibility points do not show a good agreement especially at long baselines where the 
 expected signal is much lower than the simulation's predictions. 
 
 The AGB star (Fig.~\ref{images3}) displays a pulsating stellar atmosphere with strong asymmetric structures caused by the shocks; the convective envelope is hardly visible because of the optical thickness
 of the atmosphere even without the
 inclusion of the dust opacity tables \citep{2008A&A...483..571F}. The AGB's visibilities and closure phases show large departures from symmetry already in the first lobe, but even in this case, the long baseline data points cannot be explained.
 
 The presence of spots on the reconstructed image of VX Sgr can be
qualitatively explained by the synthetic images, even if the
distribution of the spots in our
simulations has little chance to be exactly the one we
observed with AMBER. Unfortunately, we cannot constrain the 3D
simulations in term of surface intensity contrast
using these observations because we could not determine an accurate value for
the spots' flux level (Table~\ref{fitlog}).

  \begin{figure*}
\hspace*{-4mm}
   \centering
   	\begin{tabular}{cc}
  \includegraphics[width=0.35\hsize]{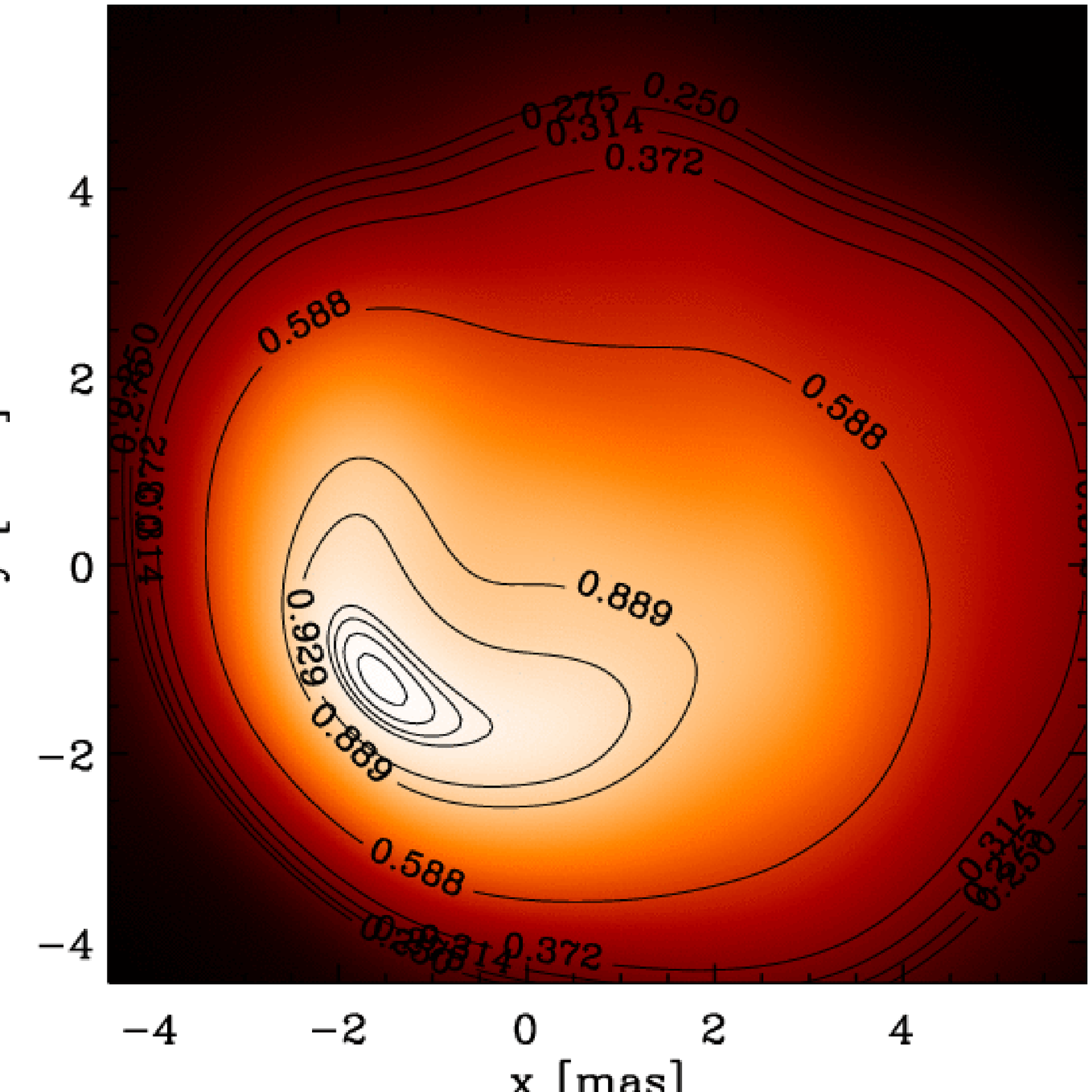}
 \includegraphics[width=0.35\hsize]{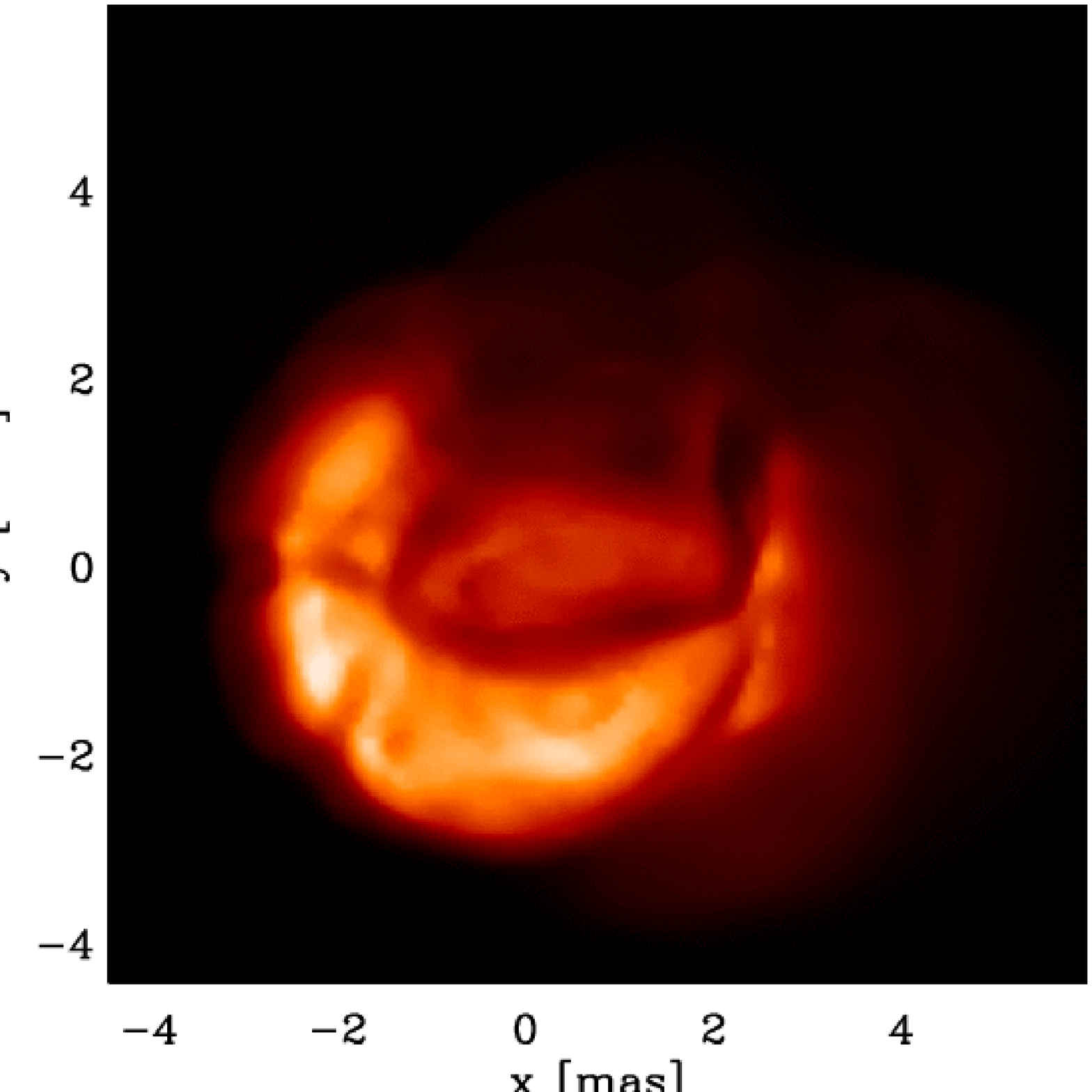}\\
 \includegraphics[width=0.4\hsize]{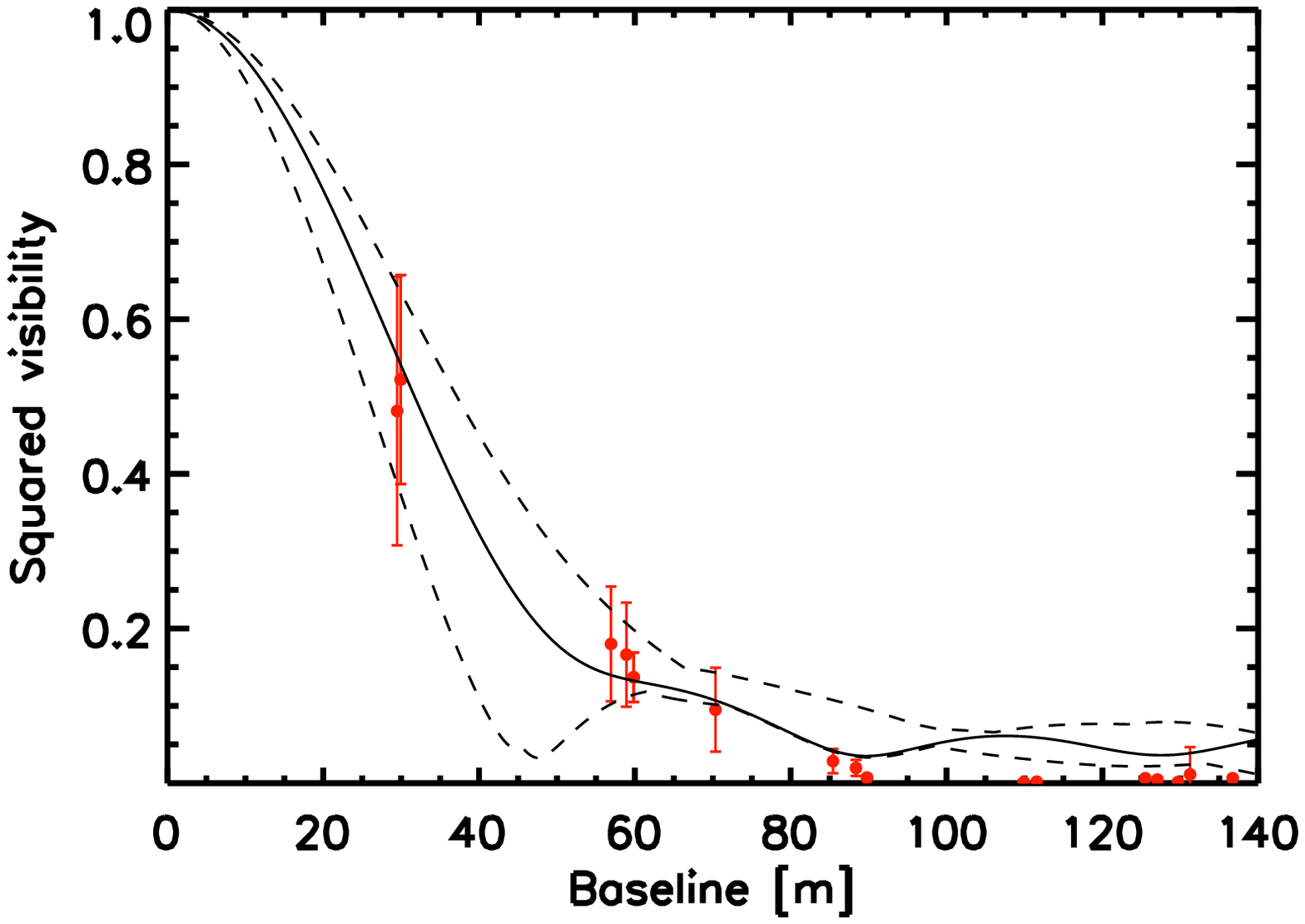}
 \includegraphics[width=0.4\hsize]{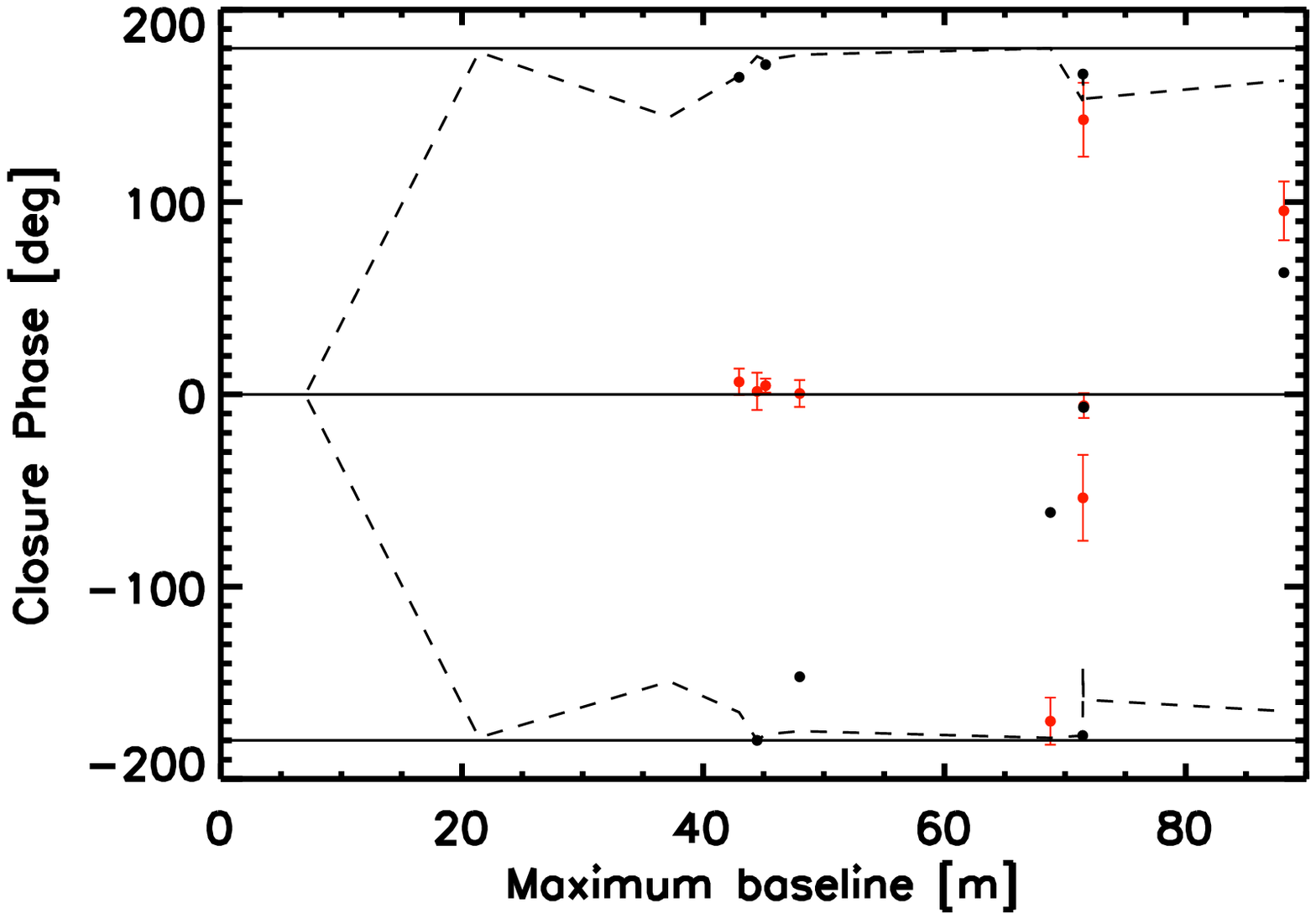}
	\end{tabular}
      \caption{Same as in Fig.~\ref{images3} for the AGB simulation \emph{Top right panel: }original AGB  image, the range is [0; $130\, 000$]\,erg\,cm$^{-2}$\,s$^{-1}$\,{\AA}$^{-1}$.
                 }
        \label{images4}
   \end{figure*}

\section{Conclusions}

Our AMBER observations unveil, for the first time, the shape of VX Sgr's
surface.

 The individual wavelength image reconstruction has been carried out
 using the MIRA software. VX Sgr displays visibilities and images with
 strong wavelength dependence: 
 (i) surface asymmetries in
 the H band ($\approx1.45-1.80$ $\mu$m) and (ii) an extended radius at $\approx2$ and
 2.35-2.50 $\mu$m. 
We find that a geometrical toy model composed of a uniform disk plus a
Gaussian disk, and three spots gives a reasonable fit for the AMBER
data. We claim that at least two spots are present on the photosphere
of VX Sgr, and we show that this is qualitatively predicted by 3D
hydrodynamical simulations of stellar convection. 
In addition, the toy model shows that one spot resides outside the
photosphere and it is brighter at longer wavelengths. The presence of this spot located outside the photosphere could be related to a complex and irregular structure in the surrounding of VX Sgr as already detected by \cite{2009A&A...504..115K} for the RSG alpha Ori. Also the two bright spots which appear at the position on the photosphere may both originate at the depth of the continuum photosphere, or higher up in the molecular layers. In fact, these spots appear at all wavelengths (Fig.~\ref{images}), also at the water bands, which hide a large part of the continuum photosphere. This might be a hint that they originate in the molecular layers far above the continuum photosphere.

We used 1D dynamical oxygen-rich Mira model predictions to explain the
visibility data points, and we found that the fit is very good at short
baselines. We conclude that H$_2$O molecules strongly affect the
visibility, and thus this molecule seems to be a dominant absorber in
the molecular layers.
The atmospheric structure of VX Sgr seems to
qualitatively resemble Mira-star models which show molecular layers
above the continuum forming layer. In addition, its photosphere shows bright 
spots that can be related to giant-cell surface convection and possibly to
the molecular layer.
However, we must point out that the Mira model used here has stellar
parameters that
are not consistent with what we expect for VX Sgr, and there are only
observations from a single epoch to do the interpretation.
Due to various uncertainties, the classification of VX Sgr must be
further investigated in the near future by, firstly, combining multi-epochs photometric and spectroscopic observations to better
determine its stellar parameters; and secondly, by observing with interferometers at higher spectral resolution to study in greater detail the changes across the H and K bands using new theoretical models with consistent stellar parameters.

\begin{acknowledgements}
We would like to thank the complete VLTI team and in particular
J.-B. Le Bouquin. Moreover, we acknowledge John Monnier, Aldo Serenelli, Josef Hron and Claudia
Paladini for the enlighten discussions. AC, BP and EJ acknowledge
financial support from ANR (ANR-06-BLAN-0105). We would like also to thank
the JMMC team. The CNRS is acknowledged for supporting us with the Guaranteed Time Observations with AMBER. We thank the variable star observations from the AAVSO International Database contributed by observers worldwide and used in this research. We thank also CINES, France, and UPPMAX, Sweden, for providing the computational resources
for the 3D simulations.
\end{acknowledgements}


\end{document}